\documentstyle[10pt,epsf,epsfig,amssymb,amsmath,multirow,array,dp_delphititle,rotating]{dp_delphi}
\makeindex
\def\DpPaperGroup{PH--EP}
\def\DpPaperRef{2009-020}
\def\DpDate{19 November 2009}
\def\DpAuthors{DELPHI Collaboration}
\def\DpSubmit{(Accepted by Eur. Phys. J. C)}
\def\DpTitle{{Measurements of $\bold{CP}$-conserving Trilinear Gauge Boson Couplings $\bold{ WWV 
(V \equiv \gamma, Z)}$ in $\bold{ e^+ e^-}$ Collisions at LEP2}}
\def\DpComment{}
\def\DpEMail{}

\newcommand{\LEP}{\mbox{\scshape Lep}}
\newcommand{\LEPII}{\mbox{\scshape Lep2}}

\newcommand{\DELPHI}{\mbox{\scshape Delphi}}

\newcommand{\ctw}{$\cos{\theta_{W^+}}$}
\newcommand{\ctl}{$\cos{\theta_l}$}
\newcommand{\clw}{$\cos{\theta_{lW}}$}
\newcommand{\wpair}{$W^+W^-$}
\newcommand{\wplus}{$W^+$}

\newcommand{\dkg}{$\Delta\kappa_\gamma$}
\newcommand{\lgamma}{$\lambda_\gamma$}
\newcommand{\dgz}{$\Delta g^Z_1$}
\newcommand{\qw}{$q_W$}
\newcommand{\muw}{$\mu_W$}
\newcommand{\qqev}{$q\bar{q}e\bar{\nu}$}
\newcommand{\qqmv}{$q\bar{q}\mu\bar{\nu}$}
\newcommand{\qqtv}{$q\bar{q}\tau\bar{\nu}$}

\newcommand{\qqvv}{$q\bar{q}\nu\bar{\nu}$}
\newcommand{\qqqq}{$q\bar{q}q\bar{q}$}

\newcommand{\qqg}{$q\bar{q}(\gamma)$}
\newcommand{\degr}{$^{\circ}$}

\newcommand{\ra}{\mbox{$\rightarrow$}}

\newcommand{\jjlv}    {\mbox{$ j j \ell \nu                                $}}
\newcommand{\jjev}    {\mbox{$ j j e \nu                                $}}
\newcommand{\jjmv}    {\mbox{$ j j \mu \nu                                $}}
\newcommand{\jjtv}    {\mbox{$ j j \tau \nu                                $}}
\newcommand{\jjjj}    {\mbox{$ j j j j                                     $}}

\newcommand{\jjX}    {\mbox{$ j j X                                        $}}
\newcommand{\lX}    {\mbox{$ \ell X                                        $}}

\newcommand{\Wev}{\mbox{${W e} \nu                                     $}}

\newcommand{\rphi}{\mbox{$r\phi$}}

\newcommand{\bq}{\begin{equation}}
\newcommand{\eq}{\end{equation}}
\newcommand{\ba}{\begin{eqnarray}}
\newcommand{\ea}{\end{eqnarray}}

\begin{document}
\makeatletter
\newcount\@tempcntc
\def\@citex[#1]#2{\if@filesw\immediate\write\@auxout{\string\citation{#2}}\fi
  \@tempcnta\z@\@tempcntb\m@ne\def\@citea{}\@cite{\@for\@citeb:=#2\do
    {\@ifundefined
       {b@\@citeb}{\@citeo\@tempcntb\m@ne\@citea\def\@citea{,}{\bf ?}\@warning
       {Citation `\@citeb' on page \thepage \space undefined}}%
    {\setbox\z@\hbox{\global\@tempcntc0\csname b@\@citeb\endcsname\relax}%
     \ifnum\@tempcntc=\z@ \@citeo\@tempcntb\m@ne
       \@citea\def\@citea{,}\hbox{\csname b@\@citeb\endcsname}%
     \else
      \advance\@tempcntb\@ne
      \ifnum\@tempcntb=\@tempcntc
      \else\advance\@tempcntb\m@ne\@citeo
      \@tempcnta\@tempcntc\@tempcntb\@tempcntc\fi\fi}}\@citeo}{#1}}
\def\@citeo{\ifnum\@tempcnta>\@tempcntb\else\@citea\def\@citea{,}%
  \ifnum\@tempcnta=\@tempcntb\the\@tempcnta\else
   {\advance\@tempcnta\@ne\ifnum\@tempcnta=\@tempcntb \else \def\@citea{--}\fi
    \advance\@tempcnta\m@ne\the\@tempcnta\@citea\the\@tempcntb}\fi\fi}
 
\makeatother

\begin{titlepage}
\pagenumbering{roman}
\CERNpreprint{\DpPaperGroup}{\DpPaperRef} 
\date{{\small\DpDate}} 
\title{\DpTitle} 
\address{\DpAuthors} 
\begin{shortabs} 
\noindent
%
\noindent
The data taken by \DELPHI{} at centre-of-mass energies between 189~and 
209~GeV are used to place limits on the $CP$-conserving trilinear gauge boson
couplings \dgz, \lgamma\ and \dkg\ associated to \wpair\ and single $W$  production at \LEPII.  Using data
from the \jjlv, \jjjj, \jjX\ and~\lX{} final states, where $j$, $\ell$ and $X$
represent a jet, a lepton and missing four-momentum, respectively,  the following limits are set on
the couplings when one parameter is allowed to vary and the others are set to their Standard Model values of zero:
\begin{center}
  \begin{tabular}{rlcrl}
               & \dgz\             & = & $-0.025^{+0.033}_{-0.030} $ & ,  \\
               & \lgamma\      & = & $ 0.002^{+0.035}_{-0.035} $ &    \\ 
   and\     & \dkg\              & = & $ 0.024^{+0.077}_{-0.081} $ & .  \\              
  \end{tabular}
\end{center}
\noindent Results are also presented when two or three parameters are allowed to vary. All observations are consistent
with the predictions of the Standard Model and supersede the previous results on these gauge coupling parameters published by \DELPHI{}.

\vspace{1cm}

{\it This paper is dedicated to the memory of Professor Paul Booth who was \DELPHI\ Deputy Spokesperson from 1989 to 1994.  He played a key role in the final installation and commissioning of the \DELPHI\ detector as well as leading the Liverpool group for many years thereafter.}
\end{shortabs}
\vfill
\begin{center}
\DpSubmit \ \\ 
\DpComment \ \\
\DpEMail \ \\
\end{center}
\vfill
\clearpage
\headsep 10.0pt
\addtolength{\textheight}{10mm}
\addtolength{\footskip}{-5mm}
\begingroup
%
\newcommand{\DpName}[2]{\hbox{#1$^{\ref{#2}}$},\hfill}
\newcommand{\DpNameTwo}[3]{\hbox{#1$^{\ref{#2},\ref{#3}}$},\hfill}
\newcommand{\DpNameThree}[4]{\hbox{#1$^{\ref{#2},\ref{#3},\ref{#4}}$},\hfill}
\newskip\Bigfill \Bigfill = 0pt plus 1000fill
\newcommand{\DpNameLast}[2]{\hbox{#1$^{\ref{#2}}$}\hspace{\Bigfill}}
%
\footnotesize
\noindent
\DpName{J.Abdallah}{LPNHE}
\DpName{P.Abreu}{LIP}
\DpName{W.Adam}{VIENNA}
\DpName{P.Adzic}{DEMOKRITOS}
\DpName{T.Albrecht}{KARLSRUHE}
\DpName{R.Alemany-Fernandez}{CERN}
\DpName{T.Allmendinger}{KARLSRUHE}
\DpName{P.P.Allport}{LIVERPOOL}
\DpName{U.Amaldi}{MILANO2}
\DpName{N.Amapane}{TORINO}
% ...
\normalsize
\endgroup

\newpage
\titlefoot{Department of Physics and Astronomy, Iowa State
     University, Ames IA 50011-3160, USA
    \label{AMES}}
\titlefoot{Physics Department, Universiteit Antwerpen,
     Universiteitsplein 1, B-2610 Antwerpen, Belgium
    \label{ANTWERP}}
\titlefoot{IIHE, ULB-VUB,
     Pleinlaan 2, B-1050 Brussels, Belgium
    \label{BRUSSELS}}
\titlefoot{Physics Laboratory, University of Athens, Solonos Str.
     104, GR-10680 Athens, Greece
    \label{ATHENS}}
\titlefoot{Department of Physics, University of Bergen,
     All\'egaten 55, NO-5007 Bergen, Norway
    \label{BERGEN}}
\titlefoot{Dipartimento di Fisica, Universit\`a di Bologna and INFN,
     Viale C. Berti Pichat 6/2, IT-40127 Bologna, Italy
    \label{BOLOGNA}}
\titlefoot{Centro Brasileiro de Pesquisas F\'{\i}sicas, rua Xavier Sigaud 150,
     BR-22290 Rio de Janeiro, Brazil
    \label{BRASIL-CBPF}}
\titlefoot{Inst. de F\'{\i}sica, Univ. Estadual do Rio de Janeiro,
     rua S\~{a}o Francisco Xavier 524, Rio de Janeiro, Brazil
    \label{BRASIL-IFUERJ}}
\titlefoot{Coll\`ege de France, Lab. de Physique Corpusculaire, IN2P3-CNRS,
     FR-75231 Paris Cedex 05, France
    \label{CDF}}
\titlefoot{CERN, CH-1211 Geneva 23, Switzerland
    \label{CERN}}
\titlefoot{Institut Pluridisciplinaire Hubert Curien, Universit\'e de Strasbourg,
     FR-67037 Strasbourg Cedex 2, France
    \label{CRN}}
\titlefoot{Now at DESY-Zeuthen, Platanenallee 6, D-15735 Zeuthen, Germany
    \label{DESY}}
\titlefoot{Institute of Nuclear Physics, N.C.S.R. Demokritos,
     P.O. Box 60228, GR-15310 Athens, Greece
    \label{DEMOKRITOS}}
\titlefoot{FZU, Inst. of Phys. of the C.A.S. High Energy Physics Division,
     Na Slovance 2, CZ-182 21, Praha 8, Czech Republic
    \label{FZU}}
\titlefoot{Dipartimento di Fisica, Universit\`a di Genova and INFN,
     Via Dodecaneso 33, IT-16146 Genova, Italy
    \label{GENOVA}}
\titlefoot{Institut des Sciences Nucl\'eaires, IN2P3-CNRS, Universit\'e
     de Grenoble 1, FR-38026 Grenoble Cedex, France
    \label{GRENOBLE}}
\titlefoot{Helsinki Institute of Physics and Department of Physical Sciences,
     P.O. Box 64, FIN-00014 University of Helsinki, 
     \indent~~Finland
    \label{HELSINKI}}
\titlefoot{Joint Institute for Nuclear Research, Dubna, Head Post
     Office, P.O. Box 79, RU-101 000 Moscow, Russian Federation
    \label{JINR}}
\titlefoot{Institut f\"ur Experimentelle Kernphysik,
     Universit\"at Karlsruhe, Postfach 6980, DE-76128 Karlsruhe,
     Germany
    \label{KARLSRUHE}}
\titlefoot{Institute of Nuclear Physics PAN,Ul. Radzikowskiego 152,
     PL-31142 Krakow, Poland
    \label{KRAKOW1}}
\titlefoot{Faculty of Physics and Nuclear Techniques, University of Mining
     and Metallurgy, PL-30055 Krakow, Poland
    \label{KRAKOW2}}
\titlefoot{LAL, Univ Paris-Sud, CNRS/IN2P3, Orsay, France
    \label{LAL}}
\titlefoot{School of Physics and Chemistry, University of Lancaster,
     Lancaster LA1 4YB, UK
    \label{LANCASTER}}
\titlefoot{LIP, IST, FCUL - Av. Elias Garcia, 14-$1^{o}$,
     PT-1000 Lisboa Codex, Portugal
    \label{LIP}}
\titlefoot{Department of Physics, University of Liverpool, P.O.
     Box 147, Liverpool L69 3BX, UK
    \label{LIVERPOOL}}
\titlefoot{Dept. of Physics and Astronomy, Kelvin Building,
     University of Glasgow, Glasgow G12 8QQ, UK
    \label{GLASGOW}}
\titlefoot{LPNHE, IN2P3-CNRS, Univ.~Paris VI et VII, Tour 33 (RdC),
     4 place Jussieu, FR-75252 Paris Cedex 05, France
    \label{LPNHE}}
\titlefoot{Department of Physics, University of Lund,
     S\"olvegatan 14, SE-223 63 Lund, Sweden
    \label{LUND}}
\titlefoot{Universit\'e Claude Bernard de Lyon, IPNL, IN2P3-CNRS,
     FR-69622 Villeurbanne Cedex, France
    \label{LYON}}
\titlefoot{Dipartimento di Fisica, Universit\`a di Milano and INFN-MILANO,
     Via Celoria 16, IT-20133 Milan, Italy
    \label{MILANO}}
\titlefoot{Dipartimento di Fisica, Univ. di Milano-Bicocca and
     INFN-MILANO, Piazza della Scienza 3, IT-20126 Milan, Italy
    \label{MILANO2}}
\titlefoot{IPNP of MFF, Charles Univ., Areal MFF,
     V Holesovickach 2, CZ-180 00, Praha 8, Czech Republic
    \label{NC}}
\titlefoot{NIKHEF, Postbus 41882, NL-1009 DB
     Amsterdam, The Netherlands
    \label{NIKHEF}}
\titlefoot{National Technical University, Physics Department,
     Zografou Campus, GR-15773 Athens, Greece
    \label{NTU-ATHENS}}
\titlefoot{Physics Department, University of Oslo, Blindern,
     NO-0316 Oslo, Norway
    \label{OSLO}}
\titlefoot{Dpto. Fisica, Univ. Oviedo, Avda. Calvo Sotelo
     s/n, ES-33007 Oviedo, Spain
    \label{OVIEDO}}
\titlefoot{Department of Physics, University of Oxford,
     Keble Road, Oxford OX1 3RH, UK
    \label{OXFORD}}
\titlefoot{Dipartimento di Fisica, Universit\`a di Padova and
     INFN, Via Marzolo 8, IT-35131 Padua, Italy
    \label{PADOVA}}
\titlefoot{Rutherford Appleton Laboratory, Chilton, Didcot
     OX11 OQX, UK
    \label{RAL}}
\titlefoot{Dipartimento di Fisica, Universit\`a di Roma II and
     INFN, Tor Vergata, IT-00173 Rome, Italy
    \label{ROMA2}}
\titlefoot{Dipartimento di Fisica, Universit\`a di Roma III and
     INFN, Via della Vasca Navale 84, IT-00146 Rome, Italy
    \label{ROMA3}}
\titlefoot{DAPNIA/Service de Physique des Particules,
     CEA-Saclay, FR-91191 Gif-sur-Yvette Cedex, France
    \label{SACLAY}}
\titlefoot{Instituto de Fisica de Cantabria (CSIC-UC), Avda.
     los Castros s/n, ES-39006 Santander, Spain
    \label{SANTANDER}}
\titlefoot{Inst. for High Energy Physics, Serpukov
     P.O. Box 35, Protvino, (Moscow Region), Russian Federation
    \label{SERPUKHOV}}
\titlefoot{J. Stefan Institute, Jamova 39, SI-1000 Ljubljana, Slovenia
    \label{SLOVENIJA1}}
\titlefoot{Laboratory for Astroparticle Physics,
     University of Nova Gorica, Kostanjeviska 16a, SI-5000 Nova Gorica, Slovenia
    \label{SLOVENIJA2}}
\titlefoot{Department of Physics, University of Ljubljana,
     SI-1000 Ljubljana, Slovenia
    \label{SLOVENIJA3}}
\titlefoot{Fysikum, Stockholm University,
     Box 6730, SE-113 85 Stockholm, Sweden
    \label{STOCKHOLM}}
\titlefoot{Dipartimento di Fisica Sperimentale, Universit\`a di
     Torino and INFN, Via P. Giuria 1, IT-10125 Turin, Italy
    \label{TORINO}}
\titlefoot{INFN,Sezione di Torino and Dipartimento di Fisica Teorica,
     Universit\`a di Torino, Via Giuria 1,
     IT-10125 Turin, Italy
    \label{TORINOTH}}
\titlefoot{Dipartimento di Fisica, Universit\`a di Trieste and
     INFN, Via A. Valerio 2, IT-34127 Trieste, Italy
    \label{TRIESTE}}
\titlefoot{Istituto di Fisica, Universit\`a di Udine and INFN,
     IT-33100 Udine, Italy
    \label{UDINE}}
\titlefoot{Univ. Federal do Rio de Janeiro, C.P. 68528
     Cidade Univ., Ilha do Fund\~ao
     BR-21945-970 Rio de Janeiro, Brazil
    \label{UFRJ}}
\titlefoot{Department of Radiation Sciences, University of
     Uppsala, P.O. Box 535, SE-751 21 Uppsala, Sweden
    \label{UPPSALA}}
\titlefoot{IFIC, Valencia-CSIC, and D.F.A.M.N., U. de Valencia,
     Avda. Dr. Moliner 50, ES-46100 Burjassot (Valencia), Spain
    \label{VALENCIA}}
\titlefoot{Institut f\"ur Hochenergiephysik, \"Osterr. Akad.
     d. Wissensch., Nikolsdorfergasse 18, AT-1050 Vienna, Austria
    \label{VIENNA}}
\titlefoot{Inst. Nuclear Studies and University of Warsaw, Ul.
     Hoza 69, PL-00681 Warsaw, Poland
    \label{WARSZAWA}}
\titlefoot{Now at University of Warwick, Coventry CV4 7AL, UK
    \label{WARWICK}}
\titlefoot{Fachbereich Physik, University of Wuppertal, Postfach
     100 127, DE-42097 Wuppertal, Germany 
    \label{WUPPERTAL}}
\titlefoot{Now at Physik-Institut der Universit\"at Z\"urich, Z\"urich, 
     Switzerland \\
\noindent
{$^\dagger$~deceased}
    \label{ZURICH}}
\nopagebreak
\clearpage
\headsep 30.0pt
\end{titlepage}
%
\pagenumbering{arabic} 
\setcounter{footnote}{0} %
\large
\section{Introduction}
\label{intro}

The reactions $e^+e^-\rightarrow W^+W^-$ and $e^+e^-\rightarrow$~\Wev\ can be used to test the non-Abelian nature of the Standard Model (SM) by studying the  trilinear couplings of the electroweak bosons~\cite{YELTGC}. In this paper, data from the final states \jjlv, \jjjj, \jjX{} and \lX{} (where $j$ represents a quark jet, $\ell$ an identified lepton and $X$ missing four-momentum) taken by \DELPHI{} at centre-of-mass energies from 189~to 209~GeV are used to determine the values of the coupling parameters which describe the trilinear $WWZ$ and  $WW\gamma$ interactions. 

The $WWV$ vertex ($V \equiv Z$~or~$\gamma$) can be described by an effective Lagrangian with 14 parameters~\cite{YELTGC,HAGIWARA}.  The set of parameters is reduced to five by assuming electromagnetic gauge invariance and by restricting the contributions in the effective Lagrangian to operators which are $C, P$-conserving. A further reduction is then achieved by extracting from the $CP$-conserving Lagrangian those terms which satisfy $\mathrm{SU(2)\otimes U(1)}$ gauge invariance, are not constrained by existing low-energy data, and are of lowest dimension ($\mathrm{\le 6}$).  This leads to a set of three independent parameters, which are studied by \DELPHI\ in the present paper: \dgz, the difference between the overall $WWZ$ coupling and its SM  value, \dkg, the deviation of the dipole coupling $\kappa_\gamma$ from its SM value, and the quadrupole coupling, \lgamma.  The imposition of gauge invariance implies relations between the dipole couplings $\kappa_\gamma$ and $\kappa_Z$ and between the quadrupole couplings $\lambda_\gamma$ and $\lambda_Z$, namely: 
$\Delta\kappa_Z = \Delta g^Z_1 - \frac{\sin^2\theta_W}{\cos^2\theta_W}\Delta\kappa_{\gamma}$ and $\lambda_Z = \lambda_{\gamma}$, 
where $\theta_W$ is the electroweak mixing angle. The terms in the effective Lagrangian which conserve $CP$, as well as $C$ and $P$ separately, correspond to the lowest order terms in a multipole expansion of $W$-$\gamma$ interactions:

\begin{align}
\label{eqn_multipole}
Q_W &=eg^{\gamma}_1 \, ,\\
\label{eqn_muw}
\mu_W &=\frac{e}{2m_W}(g^{\gamma}_1+\kappa_{\gamma}+\lambda_{\gamma}) \\
\label{eqn_qw} 
{\mathrm and} \ \ q_W &=-\frac{e}{m^2_W}(\kappa_{\gamma}-\lambda_{\gamma}) \, ,
\end{align}

\noindent where $Q_W$, $\mu_W$, and $q_W$ are respectively the charge, the magnetic dipole moment, and the electric quadrupole moment  of the $W^+$. It may be noted that electromagnetic gauge invariance, invoked above, implies the value $g^{\gamma}_1 = 1$ in these relations.

\begin{figure}[!htbp]
\begin{minipage}{0.3\textwidth}
\centering\epsfig{file=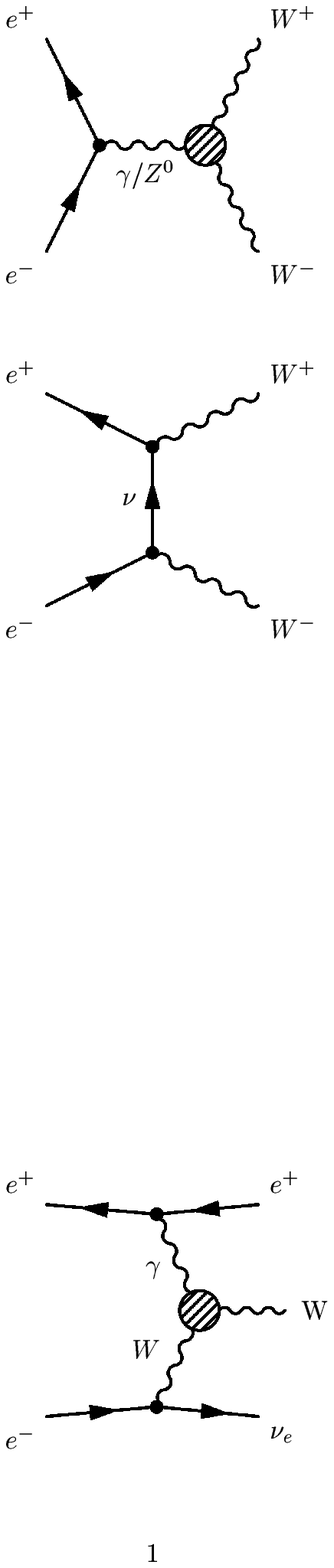,bbllx=245,bblly=618,bburx=370,bbury=725,height=0.75\textwidth,clip}
{(a)}
\end{minipage}\hfill
\begin{minipage}{0.3\textwidth}
\centering\epsfig{file=figures/WWV.ps,bbllx=245,bblly=485,bburx=370,bbury=592,height=0.75\textwidth,clip}
{(b)}
\end{minipage}\hfill
\begin{minipage}{0.3\textwidth}
\centering\epsfig{file=figures/WWV.ps,bbllx=245,bblly=180,bburx=380,bbury=287,height=0.75\textwidth,clip}
{(c)}
\end{minipage}
\caption{Diagrams contributing to \wpair~and $We\nu$ production at \LEPII.
(a) and (b) are the diagrams which describe \wpair{} production
and (c) describes $We\nu$ production.  The trilinear gauge boson vertices are denoted by
shaded circles.}
\label{fig_tgc}
\end{figure}

The diagrams which contribute to \wpair{} production are shown in figures~\ref{fig_tgc}(a) and~(b).  The $WWV$ vertex only occurs via the $s$-channel diagram shown in figure~\ref{fig_tgc}(a) and not in the $t$-channel diagram, shown in figure~\ref{fig_tgc}(b), which leads to the same final states. This reaction is studied in this paper in the final states where one $W$ boson decays to hadrons and the other decays into leptons, \jjlv, and when both $W$ bosons decay into hadrons, \jjjj.  The $WW\gamma$ vertex alone is also accessible at \LEPII{} through single $W$ production and is shown in figure~\ref{fig_tgc}(c).  This process contributes significantly in the kinematic region where the final state electron is emitted at a small angle and is studied here in two final state topologies: \lX, where the $W$ boson decays into a lepton and a neutrino, and \jjX,  where the $W$ decays into a pair of quarks.

\DELPHI\ has previously published results on charged trilinear gauge coupling parameters using data from $WW$ and \Wev\ production at energies up to 189~GeV~\cite{DELPHI_TGC172,DELPHI_TGC183,DELPHI_TGC189}, and a spin density matrix analysis of \DELPHI\ data from the \jjev\ and \jjmv\ final states at energies up to 209~GeV has been used to determine both $CP$-conserving and $CP$-violating couplings~\cite{DELPHI_SDM}. The results presented here supersede all those on $CP$-conserving couplings in these publications. Results at energies up to 209~GeV from the other \LEP\ collaborations can be found in references~\cite{RESULTS_ALEPH,RESULTS_L3,RES_SINGLEW_L3,RESULTS_OPAL}.

The \DELPHI{} detector is described in section~\ref{sec_delphi}.  The data
and simulation samples are described in section~\ref{sec_data}, the event
selection is discussed in section~\ref{sec_selection} and
section~\ref{sec_analysis} describes the analysis techniques used in the
extraction of the couplings from the data.  The different sources of
systematic uncertainties are discussed in section~\ref{sec_systematics}
and the results from fits to the data are given in section~\ref{sec_results}. 
The conclusions are presented in section~\ref{sec_conclusions}.

\section{The DELPHI detector}
\label{sec_delphi}

The \DELPHI{} detector and its performance are described in detail in~\cite{DELPHI,DELPHI_PERFORMANCE}.  For  \LEPII{} operation a number of changes were made to the sub-detectors, the trigger~\cite{TRIGGER}, the run control system and the track reconstruction algorithms to improve the performance.  The angular coverage of the Vertex Detector was extended~\cite{SITRACKER} to cover polar angles\footnote{The \DELPHI\ coordinate system has $z$~axis in the direction of the incoming $e^-$ beam. The polar angle $\theta$ is defined with respect to this direction, and the $\rphi$ plane is perpendicular to the $z$~axis.} in the range $11^\circ < \theta < 169^\circ$ with the inclusion of the Very Forward Tracker.  Together with improved tracking algorithms, alignment and calibration procedures, this resulted in an increased track reconstruction efficiency in the forward region of \DELPHI.

During the final year of operation, one sector of the twelve that constituted the central tracking device (TPC) ceased to function.  This affected around a quarter of the data collected in 2000.  The tracking algorithms were modified in this sector so as to reconstruct tracks from the signals in the other tracking detectors.  

\section{Data samples}
\label{sec_data}

A total integrated luminosity of around 600~pb$^{-1}$ was collected by \DELPHI{} between 1998 and 2000.  Table~\ref{tab_lumi} shows the
integrated luminosity available at each energy and the luminosity-weighted centre-of-mass energies.  The luminosity was determined from Bhabha scattering measurements~\cite{LUMI}.

\begin{table}[!htbp]
  \begin{center}
    \begin{tabular}{ccc}
      \hline
      Luminosity-weighted $\sqrt{s}$ (GeV)  \vline & Hadronic $L$
      (pb$^{-1}$) \vline & Leptonic $L$ (pb$^{-1}$) \\
      \hline
      \hline
      188.63 & 154.4 & 153.8 \\
      191.58 & \ 25.2 &  \ 24.5 \\
      195.51 &  \ 76.1 &  \ 72.0 \\
      199.51 &  \ 82.8 &  \ 81.8 \\
      201.64 &  \ 40.3 &  \ 39.7 \\
      204.81 &  \ 82.6 &  \ 74.9 \\
      206.55 & 135.8 & 123.7 \\
      \hline
    \end{tabular}
  \end{center}
  \caption{The centre-of-mass energies weighted by the integrated luminosity
    ($L$) for each of the \LEPII~data taking periods.  The hadronic
    luminosity was used in the fully hadronic selection and the leptonic
    luminosity was used in the other channels.  The different luminosities
    are due to tighter requirements being made on the detectors used for lepton
    identification in the semi-leptonic channels.}
  \label{tab_lumi}
\end{table}

All four-fermion final states were generated with the four-fermion generator WPHACT~\cite{WPHACT1,WPHACT2}, set up as described in~\cite{WPHAC_SETUP}.  The most recent radiative corrections to the $W$ pair production cross-section, calculated in the so-called 
Double Pole Approximation (DPA), were included via an interface to YFSWW~\cite{YFSWW}. 

The background from two-fermion production was simulated using KK2f~\cite{KK2f} and KoralZ~\cite{KoralZ}.  Additional background contributions from two-photon production were generated using BDK~\cite{BDK} and BDKRC~\cite{BDKRC}.  All of the generators were interfaced to the PYTHIA~\cite{PYTHIA1,PYTHIA2} hadronisation model tuned to the \DELPHI\ data collected at the $Z$ resonance~\cite{TUNING}.

The large simulated samples (about 1M charged current four-fermion events, 500K neutral current four-fermion events and 1M two-fermion events at each energy) were interfaced to the full \DELPHI{} simulation program DELSIM~\cite{DELPHI,DELPHI_PERFORMANCE} and passed through the same reconstruction chain as the experimental data. In order to allow analysis of data taken during the period when one part of the TPC was inoperative (as described in section~\ref{sec_delphi} above), additional samples were generated with the detector simulation modified to model this situation.

\section{Event selection}
\label{sec_selection}

In this section, the selection of events in the various final state topologies used in the determination of the coupling parameters is described.

Events selected for analysis from the $WW$ final state came from the semi-leptonic channel, \jjlv, and from the fully hadronic channel, \jjjj. The semi-leptonic final state was divided into three further channels, \jjev, \jjmv{} and \jjtv;  in the case of \jjtv\ production, only events with the tau decaying into a single charged track were considered. Events in the semi-leptonic final state are therefore characterised by two or more hadronic jets, an isolated lepton -- this comes directly from the decay of the $W$ or from the cascade decay of the tau lepton -- or a low multiplicity jet due to a hadronic tau decay, and missing momentum from the neutrino(s). The main backgrounds come from \qqg\ production and from four-fermion final states of two quarks and two leptons of the same flavour. The analysis of the fully hadronic final state from $WW$ production involved a search for four-jet events in which the di-jet invariant masses of one of the pairings into two di-jets were compatible with the $W$ mass. Here, also, \qqg\ production represents a major source of background, with some contamination also from $ZZ$ decays into \qqqq\ and $q{\bar q}\tau^+\tau^-$.

Events from single $W$ production, \Wev,  were selected in the kinematic region where the final state electron is very close to the beam direction and remains undetected. Final states with hadronic $W$ decays and with leptonic decays into electron or muon and a neutrino were considered, so that the topologies analysed were \jjX\ and \lX, with $l \equiv e, \mu$ and $X$ representing missing momentum. The main background contributing to the \jjX\ topology came from \qqtv\ production. In the $eX$ channel the major source of background was from $e^+e^-\gamma (\gamma)$ production with one electron (or positron) and the final state photon(s) unobserved, while in the $\mu X$ topology the main backgrounds were from $ee\mu\mu$ production, mainly via two-photon processes, and from $\mu\mu\gamma$ production. Some of these background processes (such as \qqtv\ production) themselves contain triple gauge boson vertices in their production mechanisms, and thus contribute to the precision of the results. 

Full details of the event reconstruction procedure adopted, and of the selection of events in the channels considered here from $WW$ production can be found in~\cite{WWXSEC}, \DELPHI's report on the measurement of the $WW$ production cross-section, while the selection procedure for events in the \Wev\ final state is very similar to that used in our previous publication of charged trilinear gauge boson couplings at 189~GeV~\cite{DELPHI_TGC189}. In the following sections, a summary of these procedures is given.

The total numbers of events selected at each centre-of-mass energy are given in table~\ref{tab_events}. The table also gives examples (at 200~GeV) of the event selection efficiencies and estimated background cross-sections; the errors on these cross-sections are treated as a systematic uncertainty and are discussed in section~\ref{sec_systematics}. 

\subsection{Particle selection}
 
Reconstructed charged particles were required to have momentum greater than 0.1~GeV/$c$ and less than 1.5~times the beam momentum, a relative momentum error less than~1, an impact parameter in $\rphi$ less than 4~cm, and a $z$ impact parameter less than 4~cm/$\sin\theta$.  Neutral clusters were required to have energy exceeding 300~MeV in the barrel electromagnetic calorimeter (HPC) and exceeding 400~MeV and 300~MeV in the two forward electromagnetic calorimeters (FEMC and STIC, respectively). Electron identification was based on the association of energy deposits in the electromagnetic calorimeters with momentum measurements in the tracking chambers (the Inner Detector and the TPC) and, in the case of lower energy candidates, with energy loss measurements in the TPC. Muon candidates were identified by extrapolating tracks through the entire detector and associating them with energy deposits recorded in the hadron calorimeter (HCAL) and hits recorded in the muon chambers.

\subsection{Selection of events in the \jjlv\ final state}
\label{sec_sel_jjlv}

The selection of events in the semi-leptonic final state involved cut-based selections, followed by the application of an Iterative Discriminant Analysis (IDA)~\cite{IDA1,IDA2}.

An initial hadronic pre-selection was applied where at least 5~charged particles were required, the energy of the charged particles had to be at least 10\% of the centre-of-mass energy, and the following condition was imposed: $\sqrt{EMF_f^2+EMF_b^2} < 0.9\times E_{beam}$, where $EMF_{f,b}$ are the total energy deposited in the electromagnetic calorimeters in the forward and backward directions, defined as two cones around the beam axes of half-angle~20\degr\ .

At this point a search was made for leptons, allowing each event to have up to three lepton candidates, one of each flavour. Of the electrons found in the particle selection procedure, the one with the highest value of $E \times \theta_{iso}$ was chosen as the electron candidate. Here $E$ is the measured electron energy and $\theta_{iso}$ is the isolation angle of the electron track, defined as the angle made to the closest charged particle with momentum greater than 1~GeV/$c$. The candidate was then required to have energy greater than 15~GeV. Similarly, an identified muon track with momentum~$p$ was selected if it had the highest value of $p \times \theta_{iso}$ and if its momentum exceeded 15~GeV/$c$. The event was then clustered into jets using LUCLUS~\cite{PYTHIA1,PYTHIA2} with $d_{join}$ = 6.5~GeV/$c$.  Particles were removed from the jets if they were at an angle greater than~20\degr\ to the highest energy particle and the remaining jet with the lowest momentum-weighted spread\footnote{defined as $\frac{\sum_{i} \theta_i\cdot |p_i|}{\sum_{i} |p_i|}$ where $\theta_i$ is the angle made by the momentum $p_i$ of the $i$th particle in the jet with the total jet momentum.} was considered as a tau candidate.  Particles were removed from this jet if they were at angle greater than~8\degr\ from the jet axis and the remaining jet was required to contain at least one charged particle.

For each lepton candidate, the remaining particles in the event were clustered into two jets using the DURHAM algorithm~\cite{DURHAM}.  Each of these jets was required to contain at least three particles, of which at least one had to be charged.  A further pre-selection was made before applying the full selection using the IDA: for \jjev\ and \jjmv\ candidates the transverse energy was required to be greater than 45~GeV; the missing momentum had to exceed 10~GeV/$c$; the visible energy divided by the centre-of-mass energy at which the IDA was trained (defined below), $E_{vis}/E_{train}$, was required to be between~40\% and~110\%; and the fitted $W$ mass from a constrained kinematic fit (imposing four-momentum conservation and equal mass for the two $W$ bosons in the event) had to be greater than 50~GeV/$c^2$.  For \jjtv~candidates, the transverse energy was required to be greater than 40~GeV, the missing momentum between 10~and 80~GeV/c, the ratio $E_{vis}/E_{train}$ between 35\% and 100\%, and the fitted $W$ mass greater than 50~GeV/$c^2$.

After the pre-selection cuts an extended IDA analysis was used which treated correctly the case where the signal and background had different shapes. The input observables were transformed to make their distributions Gaussian.  The IDA was trained on 50k four-fermion events for charged and neutral processes and 100k \qqg\ events at three centre-of-mass energies: 189, 200 and 206~GeV. The following variables were used in the selection of all channels: the total multiplicity, the visible energy, the lepton isolation angle, the ratio between the reconstructed effective centre-of-mass energy, $\sqrt{s'}$,~\cite{SPRIME} and the centre-of-mass energy, $\sqrt{s}$, the magnitude and the polar angle of the missing momentum, and the fitted $W$ mass. The lepton energy was used in the selection of \jjev\ and \jjmv\ candidates. The angle between the lepton and the missing momentum was used in the \jjmv\ and \jjtv\ selections.  For the \jjev\ selection, the transverse energy was also used.  In addition, the aplanarity\footnote{defined as $\frac{3}{2}\lambda_3$ where $\lambda_3$ is the smallest eigenvalue of the sphericity tensor $S^{\alpha\beta}=\frac{\sum_i{p_i^\alpha p_i^\beta}}{\sum_i{|p_i|^2}}$. The $p_i$ are the three momenta of the particles in the event and $\alpha,\beta=1,2,3$ correspond to the $x$, $y$, $z$ momentum components.}, the charged multiplicity of the tau jet and its momentum-weighted spread were used in the \jjtv~selection. The cut on the output of the IDA was chosen such that it maximised the value of the efficiency times the purity for each channel. 

In the application of the IDA,  events with more than one lepton candidate, one of which was a muon, were first passed through the \jjmv\ selection procedure; those not selected, but containing an electron candidate, were then passed to the \jjev\ selection, and if the event failed both the muon and electron selection procedures and included a tau candidate, it was passed to the \jjtv\ selection.  A final cut, requiring the charged multiplicity of the tau jet to be~1, ensured that the charge of the $W$ boson which produced it was well determined.

At centre-of-mass energy of 200~GeV, the efficiencies of the \jjev, \jjmv{} and \jjtv{} selections were found to be 71.0\%, 88.2\% and 54.6\%, respectively (see table~\ref{tab_events}). The selection efficiencies differed by no more than~2\% over the energy range considered. The respective background cross-sections for the three channels at 200~GeV were evaluated to be 0.232~pb, 0.075~pb and 0.344~pb, with the main contributions coming from \qqg\ and from neutral current four-fermion final states. Combining the data at all centre-of-mass energies, totals of 1101, 1246 and 886~events were selected in the three leptonic channels, respectively.  

\subsection{Selection of events in the fully hadronic final state}

In the selection of fully hadronic final states, the charged and neutral particles in each event 
were forced into a four-jet configuration with the DURHAM algorithm.  
A pre-selection was performed where the reconstructed effective centre-of-mass energy, $\sqrt{s'}$, 
was required to be greater than 65\% of the nominal centre-of-mass energy, the total and 
transverse energy for charged particles were each required to be greater than 20\% of the
nominal centre-of-mass energy, the total multiplicity for each jet had to exceed~3,
the condition $y_{cut}>0.0006$ was imposed for the migration of 4~jets to 3~jets
when clustering with the DURHAM algorithm, and a four-constraint kinematic fit of
the measured jet energies and directions, imposing four-momentum
conservation, was required to converge.

A feed-forward neural network, based on the JETNET package~\cite{JETNET} was
then used to improve the rejection of two- and four-fermion backgrounds. The
network uses the standard back-propagation algorithm and consists of three
layers with 13~input nodes, 7~hidden nodes and one output node.  The choice
of input variables was optimised~\cite{NNOPTIMISATION} to give the greatest
separation between $WW$ and two-fermion events.  The following jet and event
observables were used as input variables: the difference between the minimum
and maximum jet energies after the kinematic fit, the minimum angle between the jets
after the fit, the value of $y_{cut}$ from the DURHAM algorithm for the
migration of 4 jets into 3 jets, the minimum particle multiplicity of any
jet, the reconstructed effective centre-of-mass energy, the maximum
probability amongst each of the 3 possible jet pairings of a six-constraint fit (imposing the 
additional constraints that the invariant mass of each jet pair should be equal to 
the $W$ mass, set equal to 80.40~GeV/$c^2$), the thrust, the sphericity, the transverse energy, the sum of the cubes of the 
magnitudes of the momenta of the 7~highest momentum particles, 
$\sum_{i=1}^7 |\vec{p}_i|^3$, the minimum jet broadening, $B_{min}$~\cite{DURHAM}, 
and the Fox-Wolfram moments $H3$ and $H4$~\cite{FOX-WOLFRAM}.

The neural network was trained on separate samples of 2500 signal and
$Z/\gamma \rightarrow q\overline{q}$ events for each centre-of-mass energy.
The network output was calculated for other independent four-fermion, 
two-fermion and two-photon processes.

The efficiency of the fully hadronic selection for a centre-of-mass energy of 200~GeV was estimated to be 81.9\% (see table~\ref{tab_events}); the efficiency varied by no more than~4\% over the energy range considered.  The background cross-section at 200~GeV was evaluated to be 1.21~pb, with the main contribution coming from \qqg. Combining the data at all centre-of-mass energies, a total of 4348 events was selected.

\subsection{Selection of events in the \jjX\ final state}
\label{sec_sel_jjX}

Events were considered as \jjX{} candidates if there were no identified leptons with momentum greater than 12~GeV/$c$, the measured transverse momentum exceeded 20~GeV/$c$, and the invariant mass of detected particles lay between 45~GeV/$c^2$ and 90~GeV/$c^2$. 
In addition, events were rejected if any neutral clusters were found in the electromagnetic or hadronic calorimeters with energy exceeding 1~GeV within a cone of half-angle 30\degr\ around the direction of the missing momentum. Particles were clustered into jets using LUCLUS with $d_{join}$ = 6.5~GeV/$c$ and events were required to have two or three jets only.  Surviving events were forced into a two-jet configuration and accepted if the jet polar angles were between 20\degr\ and 160\degr\ and the acoplanarity angle\footnote{defined as the angle between the planes containing each jet direction and the beam direction.} between the jets was less than 160\degr. 

The efficiency of the selection is quoted with respect to a reduced phase space defined by the following generator level cuts: the acoplanarity angle between the quarks was required to be less than 170\degr; the invariant mass of the quark pair had to be greater than 40~GeV/$c^2$; the quark directions were required to have polar angles between 20\degr\ and 160\degr; and the electron polar angle was required to be less than 11\degr\ or greater than 169\degr.  The efficiency for selecting the \Wev\ final state with $W \ra\ q{\bar q}$ was found to be between 43.7\% and 48.0\%, depending on the centre-of-mass energy, with a luminosity-weighted mean value of 45.4\%; 215 events in total were selected in the data. For Standard Model values of the couplings, a total of $219.8 \pm1.6$ events was expected, comprising 79.5 events from \qqev\ production with the electron or positron lost in the beam pipe, 13.0 events from \qqev\ production with the electron or positron elsewhere in the detector, 18.5 events from \qqmv\ production, 67.0 events from \qqtv, 36.0 events from \qqvv, and 5.8 events from \qqg\ production. The error in the expected total number of events arises from the statistical errors in the selection efficiencies estimated for the contributing processes. 
All the processes contributing to the selected sample except \qqg\ production include diagrams with trilinear gauge couplings, and this was taken into account in the subsequent analysis. The background cross-section of 0.048~pb shown for the \jjX\ channel in table~\ref{tab_events} represents the contribution at $\sqrt{s} = 200$~GeV from \qqg\ production.

\subsection{Selection of events in the \lX~final state}
\label{sec_sel_lX}

To be considered as an \lX{} candidate, events were required to have only one charged particle, clearly identified as an electron or a muon from signals in the electromagnetic calorimeters or the muon chambers, respectively, using the same procedures as described in the selection of semi-leptonic events (section~\ref{sec_sel_jjlv}). The impact parameter for the lepton was required to be less than 0.1~cm in the $\rphi$ plane and less than 4~cm in the $z$ direction.  The lepton candidate was required to have momentum less than 75~GeV/$c$, with the transverse component of this momentum greater than 20~GeV/$c$.  The total energy deposited in the electromagnetic calorimeter, but not associated with the track, was required to be less than 5~GeV.  The ratio of the energy deposited by electron candidates in the electromagnetic calorimeter
to that determined from the measured value of the momentum was required to exceed 0.7. 

As in the case of the \jjX\ final state described above, the efficiency of the selection was calculated in a reduced phase space region defined by cuts made at generator level; for the \lX\ final state, these were defined as follows: the lepton energy was required to be less than 75~GeV; the transverse momentum of the lepton had to be greater than 20~GeV/c; and the polar angle of the missing momentum was required to be in the range from zero to 11\degr\ or between 169\degr\ and 180\degr. Totals of 37 and 39 candidates were selected in the $\mu X$ and $e X$ channels, respectively, with luminosity-weighted average efficiencies for selection of the \Wev\ final states of 49.4\% for $W \ra \mu\nu$ and 31.3\% for $W \ra e\nu$. For Standard Model values of the couplings, $34.0 \pm 1.4$ and $31.9 \pm 1.5$ events were expected in the two channels, respectively. The predicted $\mu X$ sample comprised 17.7 events from $e\mu\nu\bar{\nu}$ production with the electron or positron lost in the beam pipe, 1.6 events from $e\tau\nu\bar{\nu}$ production, also with an invisible electron or positron, 1.9 events from $\mu\mu\nu\bar{\nu}$ production, 2.0 events from $\mu\tau\nu\bar{\nu}$ production, 4.0 events from $\mu\mu e e$, and 6.8 events from $\mu\mu (\gamma)$. In the $e X$ sample, 19.2 events were expected from $e e \nu\bar{\nu}$ production with one lost electron or positron, 1.6 events and 0.7 events, respectively, from $e\mu\nu\bar{\nu}$ and $e\tau\nu\bar{\nu}$ with the electron or positron in the beam pipe, 1.5 events from  $e\tau\nu\bar{\nu}$ production with the electron or positron elsewhere in the detector, and 8.8 events from Compton and Bhabha scattering with only one electron (or positron) detected in the final state. The background cross-sections in the $\mu X$ and $e X$ final states at 200~GeV quoted in table~\ref{tab_events}, 0.016~pb and 0.013~pb, respectively,  represent the contributions from the processes contributing to these final states which have no dependence on the trilinear gauge couplings under consideration, namely the $\mu\mu (\gamma)$ contribution to $\mu X$ and the Compton and Bhabha contributions to $e X$. All the other contributions to these final states have a dependence on trilinear gauge couplings in their production, and this was taken into account in the subsequent analysis.

\begin{table}
  \begin{center}
    \begin{tabular}{crrrrrrr}
      \hline
      Energy (GeV) & \jjev & \jjmv & \jjtv & \jjjj & \jjX & $\mu X$ & $e X$ \\
      \hline
      \hline
      189    &  269  &  336  &  236  & 1042  &  64  & 11 & 10  \\
      192    &   42  &   53  &   37  &  187  &   4  &  1 &  1  \\
      196    &  151  &  166  &  116  &  532  &  22  &  6 &  5  \\
      200    &  162  &  190  &  145  &  614  &  24  &  6 &  6  \\
      202    &   94  &   89  &   57  &  317  &  12  &  5 &  3  \\
      205    &  169  &  153  &   94  &  657  & \multirow{2}{*}{89} &
      \multirow{2}{*}{8} & \multirow{2}{*}{14} \\
      207    &  214  &  259  &  201  &  999  &     &     &     \\
      \hline
      TOTAL  & 1101  & 1246  &   886 & 4348  & 215 & 37  & 39  \\
      \hline
      \hline
    $\epsilon$($\sqrt{s}$ = 200~GeV) (\%)           & 71.0 & 88.2 & 54.6 & 81.9 & 48.0 & 50.8 & 31.9 \\ 
    $\sigma_{back}$($\sqrt{s}$ = 200~GeV) (pb) & 0.232 & 0.075 & 0.344 & 1.21 & 0.048 & 0.016 & 0.013 \\
      \hline
    \end{tabular}
  \end{center}
  \caption{The numbers of events selected from the data in each channel for
    each centre-of-mass energy.  The selection efficiencies $\epsilon$ and background
    cross-sections $\sigma_{back}$ are shown for the centre-of-mass energy of 200~GeV.}
  \label{tab_events}
\end{table}

\section{Determination of the couplings}
\label{sec_analysis}

The extraction of the couplings from the data exploited the fact that the differential cross-section,
$\frac{d\sigma}{d\vec{\Omega}}$, is quadratic in the set of couplings, $\alpha_i (\equiv$~ \dgz, \dkg, \lgamma), and can be expressed as

\begin{equation}
\frac{d\sigma}{d\vec{\Omega}} = {c_{0}}(\vec{\Omega})
+\sum_{i}{c_{1}^{i}}(\vec{\Omega}){\alpha_{i}}
+\sum_{i\le j}{c_{2}^{ij}}(\vec{\Omega}){\alpha_{i}\alpha_{j}} \, ,
\label{eqn_OO}
\end{equation}

\noindent where $\vec{\Omega}$ represents the kinematic phase space variables and $i,j$ are summed over the 
number, $N$, of parameters being determined. The coefficients $c_1^i$ and $c_2^{ij}$ were calculated using 
WPHACT for the final states coming from $W^+W^-$  production and using DELTGC~\cite{DELTGC} for single $W$ final states.  
This allows the fully simulated events to be re-weighted to non-SM values of the couplings.

\subsection{Semi-leptonic final state}

The analysis of the data in the semi-leptonic channel used the method of Optimal Observables~\cite{OO1,OO2,OO3}, in which an expansion of the form~(\ref{eqn_OO}) represents the first two terms in a Taylor expansion of the differential cross-section for any process in terms of a set of $N$ parameters $\alpha_i$. If it is known that the $\alpha_i$ are small, then the $N$ lowest order terms in~(\ref{eqn_OO}) contain most of the information needed for the determination of the parameters. In the present case, where the amplitude for the processes we consider is linear in the parameters, the Taylor expansion is truncated at the second order, and~(\ref{eqn_OO}) gives the value of the cross-section without approximation. This suggests an analysis in terms of the quantities  

\begin{equation}
\omega_1^i(\vec{\Omega}) = \frac{c_1^i(\vec{\Omega})}{c_0(\vec{\Omega})} 
\ \ {\mathrm and} \ \ 
\omega_2^{ij}(\vec{\Omega}) = \frac{c_2^{ij}(\vec{\Omega})}{c_0(\vec{\Omega})} \, ,
\label{eqn_OV}
\end{equation}

\noindent which are easily derived from the differential cross-section. Such an analysis is described in~\cite{OOTGC}, where the probability distribution function, $P(\vec{\Omega}, \vec{\alpha})$, for observing an event at phase space position $\vec{\Omega}$ when the parameters have values $\vec{\alpha}$ ($\equiv  \alpha_1 \, ... \, \alpha_N$) is projected in the $\omega_1^i(\vec{\Omega})$ and $\omega_2^{ij}(\vec{\Omega})$ of~(\ref{eqn_OV}), the Optimal Variables.

When $\vec{\Omega}$ is known precisely, a fit to the Optimal Variables allows the couplings to be determined with a precision equal to that of an unbinned maximum likelihood fit over all of the phase space variables.  In practice, the measured values of the Optimal Variables are defined by the convolution of the differential cross-section with the resolution and efficiency functions of the detector. However, it has been confirmed by Monte Carlo tests~\cite{OOOPTIMISATION} that little loss of precision occurs when this convolution is performed. In the case where one parameter, $\alpha_i$, is free to deviate from its Standard Model value, two Optimal Variables ($\omega_1^i$ and $\omega_2^{ii}$) contain the whole information, but five (or nine) Optimal Variables are required when two (or three) parameters are released from their Standard Model values. For one-parameter fits, there is an obvious advantage in simplicity in the use of the Optimal Variable method over an analysis using the five angular variables (the $W$ production angle and the $W^+$ and $W^-$ decay angles) known to contain most of the information on the coupling parameters in $WW$ production, while in the case of multi-parameter fits the number of Optimal Variables is equal to or greater than the number of angular variables. We have compared these methods using simulated event samples: in all cases - for one-, two- and three-parameter fits - the precision obtained from the Optimal Variable analysis was at least as good as that from the angular analysis, allowing us to use the same methodology throughout the analysis.

\begin{figure}[!htbp]
\centering\epsfig{file=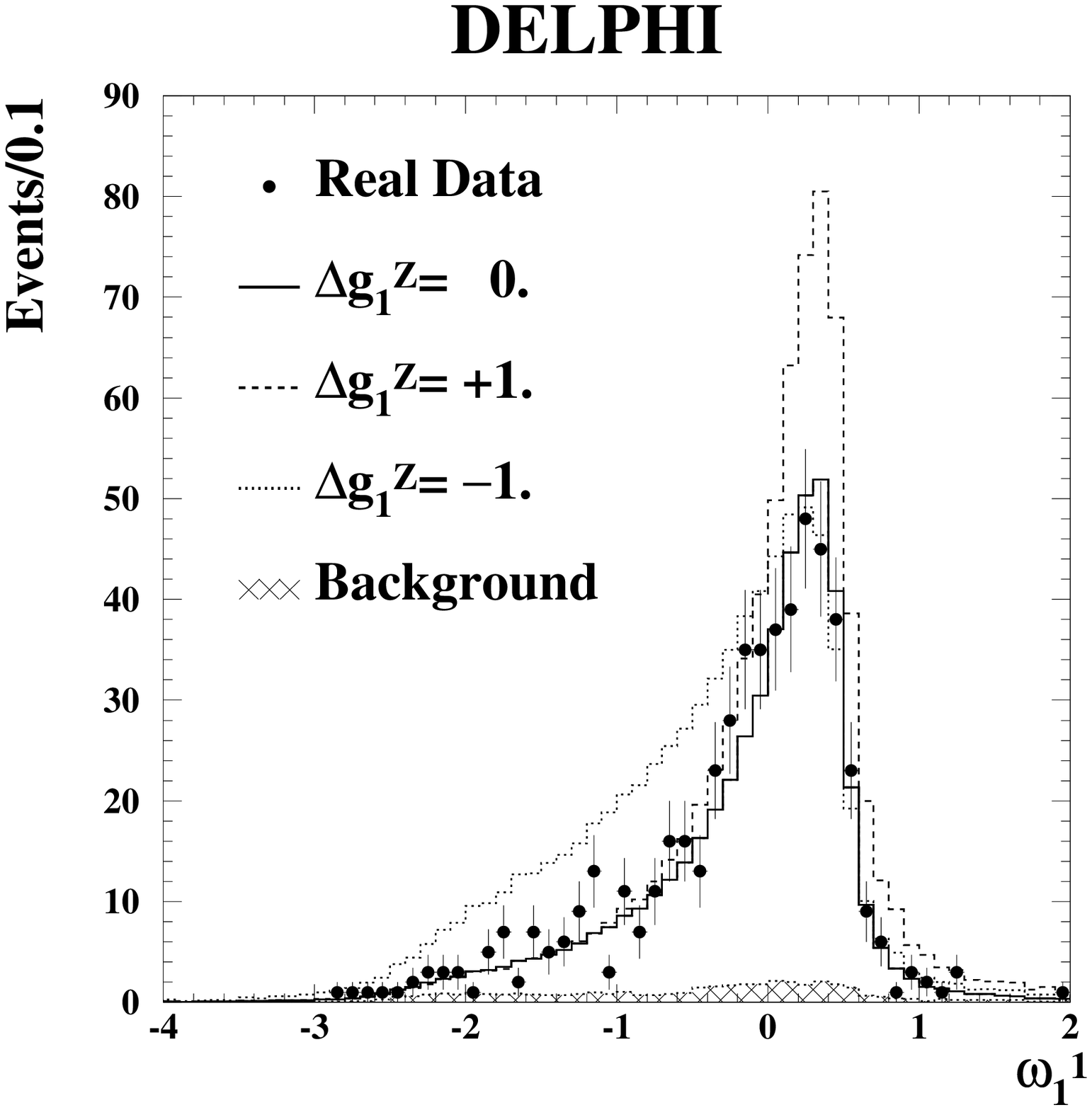,height=0.65\textwidth,bbllx=10,bblly=135,bburx=515,bbury=640,clip} \\
\centering\epsfig{file=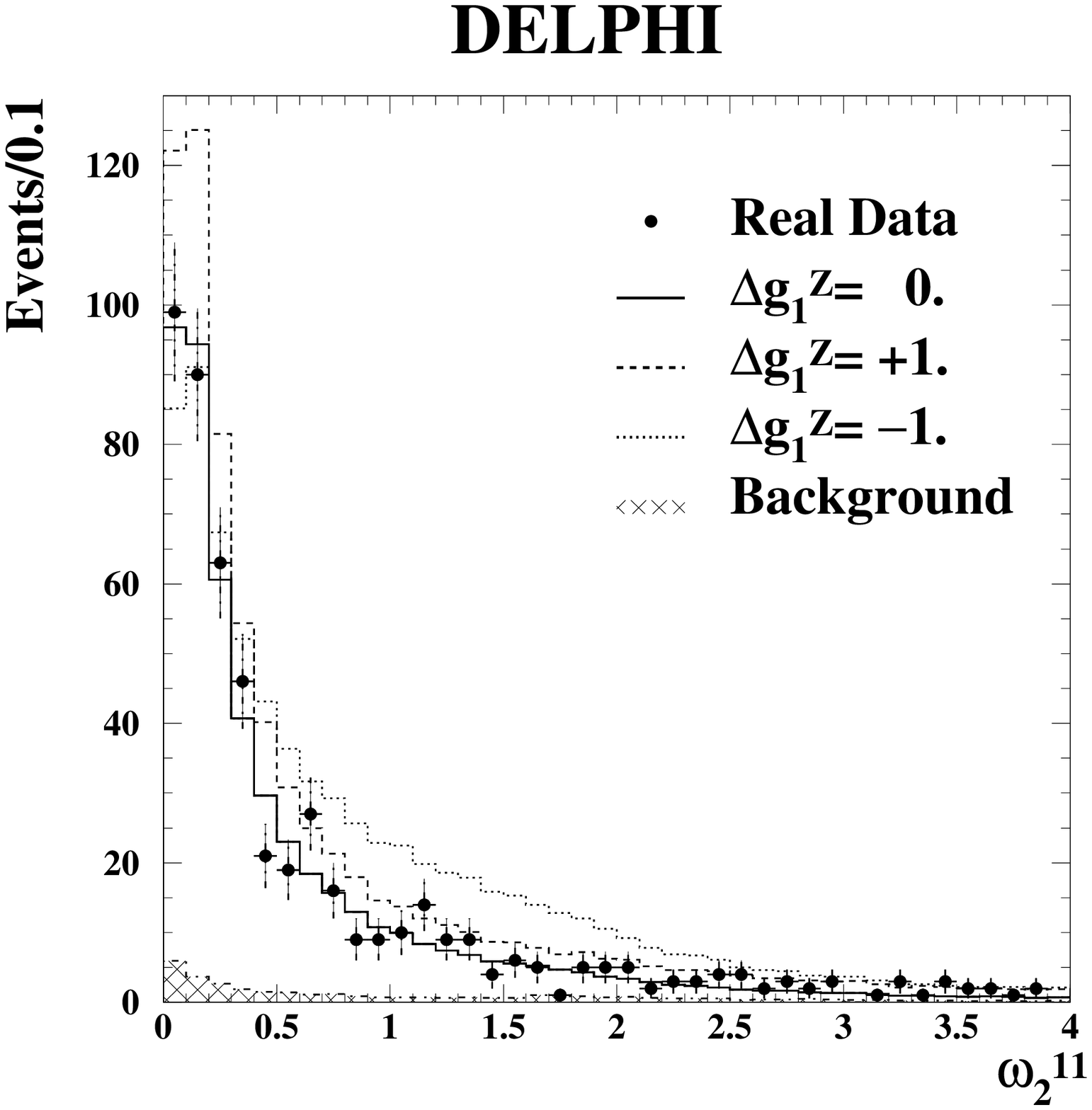,height=0.65\textwidth,bbllx=10,bblly=135,bburx=515,bbury=640,clip}
\caption
    {Distribution of the Optimal Variables $\omega_{1}^{1}$ and $\omega_{2}^{11}$ (the coefficients, respectively, of \dgz\ and of (\dgz)$^2$ in the expansion of the differential cross-section in terms of Optimal Variables) for  semi-leptonic data at 200~GeV.  The points represent the real 
data, the solid lines the expected distributions for the SM value of the coupling, and the dashed lines the expected  distributions for the non-SM values \dgz=$\pm 1$.  The shaded area represents the background. The simulated distributions are normalised to the same luminosity as the data.}
\label{fig_OO_dkg}
\end{figure}

\begin{figure}[!htbp]
\centering\epsfig{file=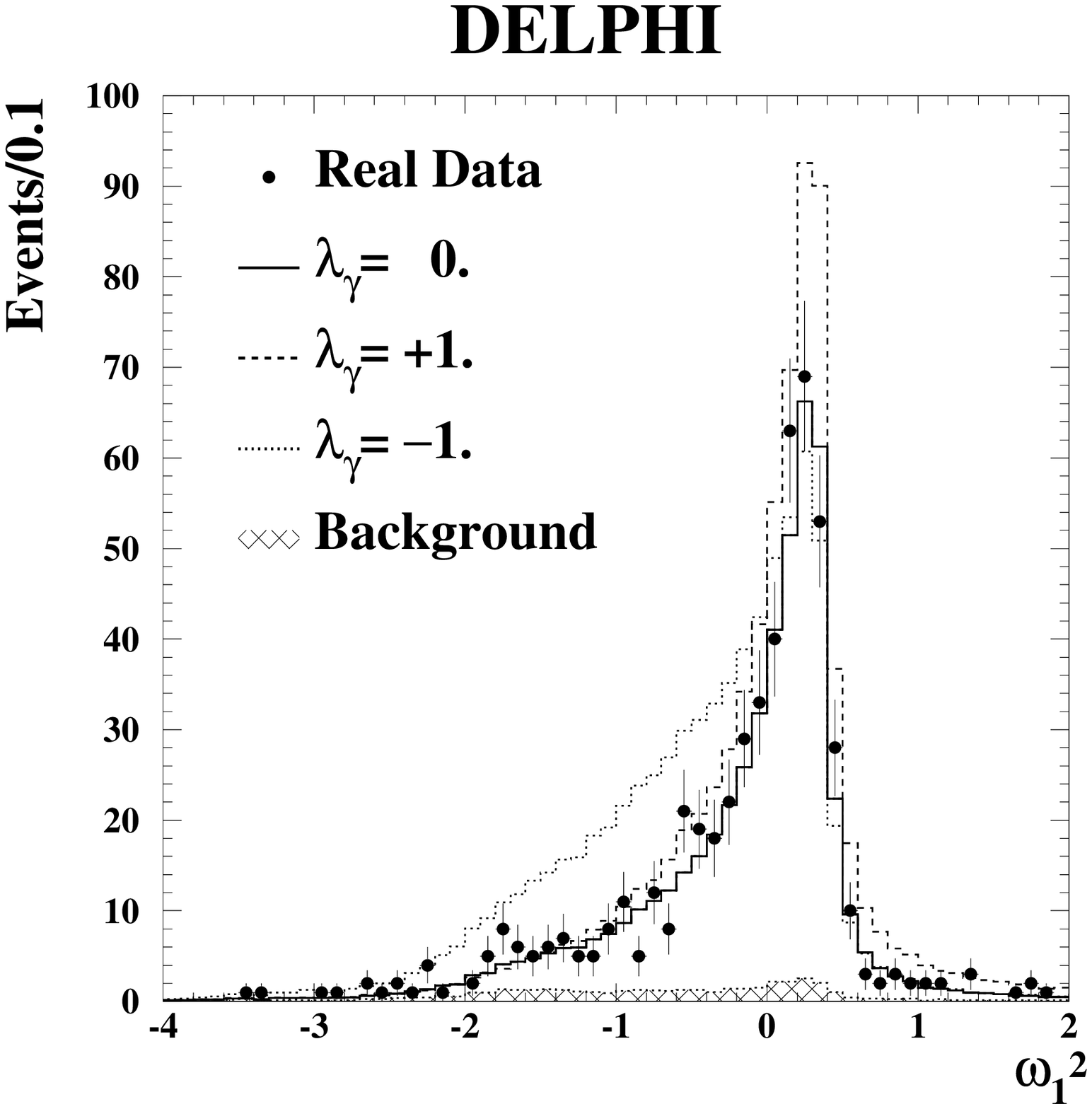,height=0.65\textwidth,bbllx=10,bblly=135,bburx=515,bbury=640,clip} \\
\centering\epsfig{file=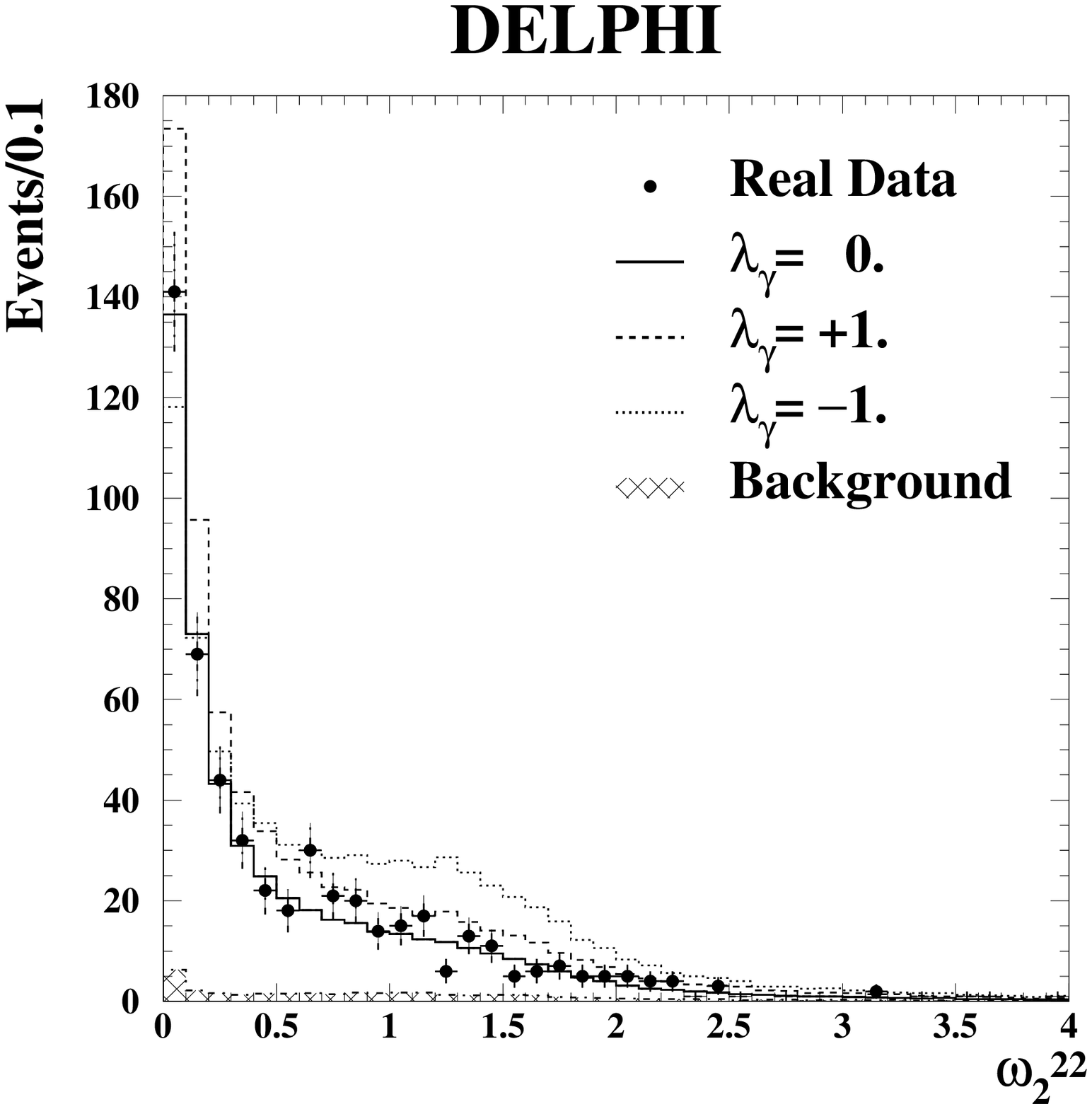,height=0.65\textwidth,bbllx=10,bblly=135,bburx=515,bbury=640,clip}
\caption
    {Distribution of the Optimal Variables $\omega_{1}^{2}$ and $\omega_{2}^{22}$ (the coefficients, respectively, of \lgamma\ and of (\lgamma)$^2$ in the expansion of the differential cross-section in terms of Optimal Variables) for semi-leptonic data at 200~GeV.  The points represent the real data, the solid lines the expected distributions for the SM value of the coupling, and the dashed lines the expected  distributions for the non-SM values \lgamma=$\pm 1$.  The shaded area represents the background. The simulated distributions are normalised to the same luminosity as the data.}
\label{fig_OO_dgz}
\end{figure}

\begin{figure}[!htbp]
\centering\epsfig{file=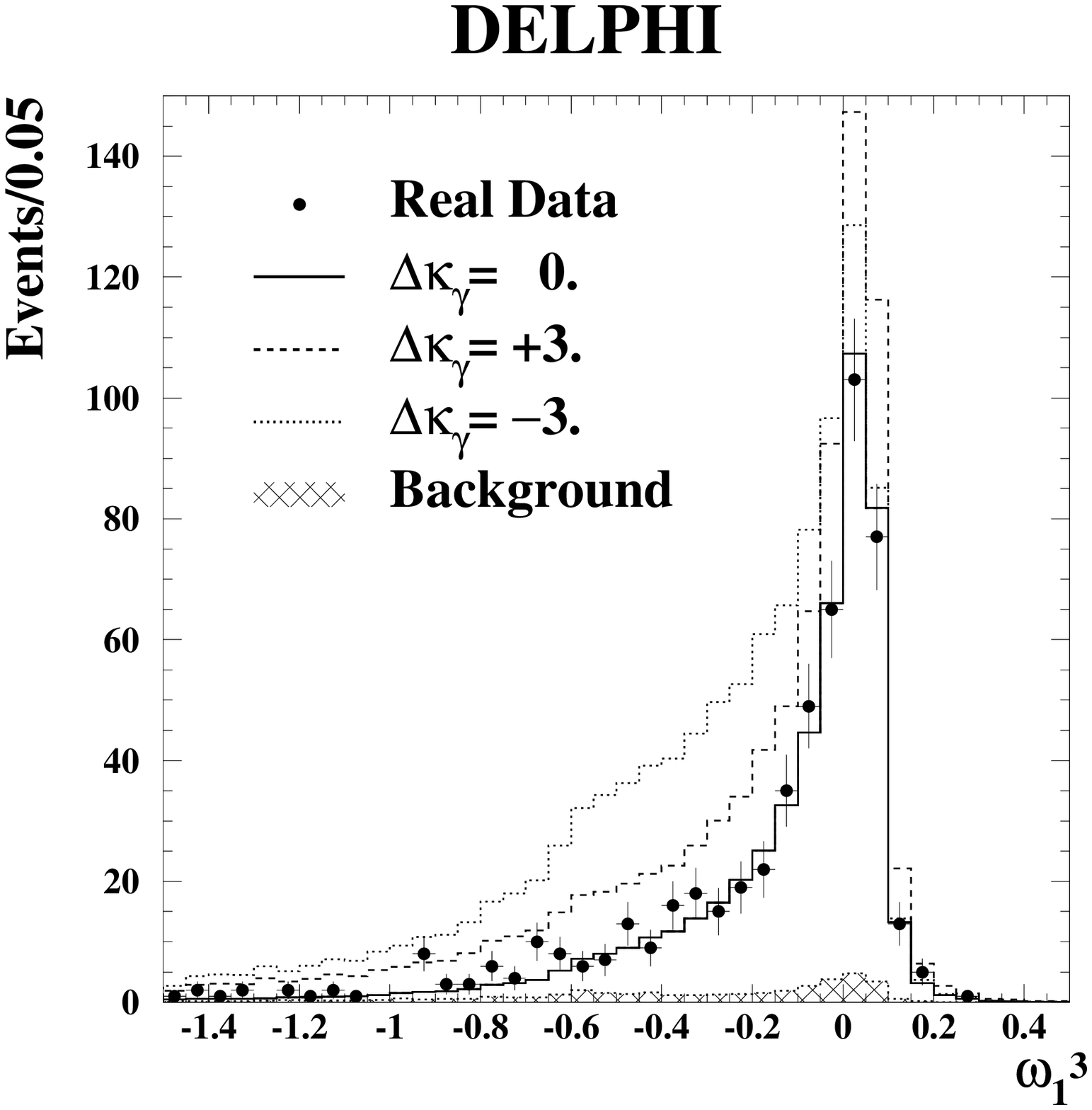,height=0.65\textwidth,bbllx=10,bblly=135,bburx=515,bbury=640,clip} \\
\centering\epsfig{file=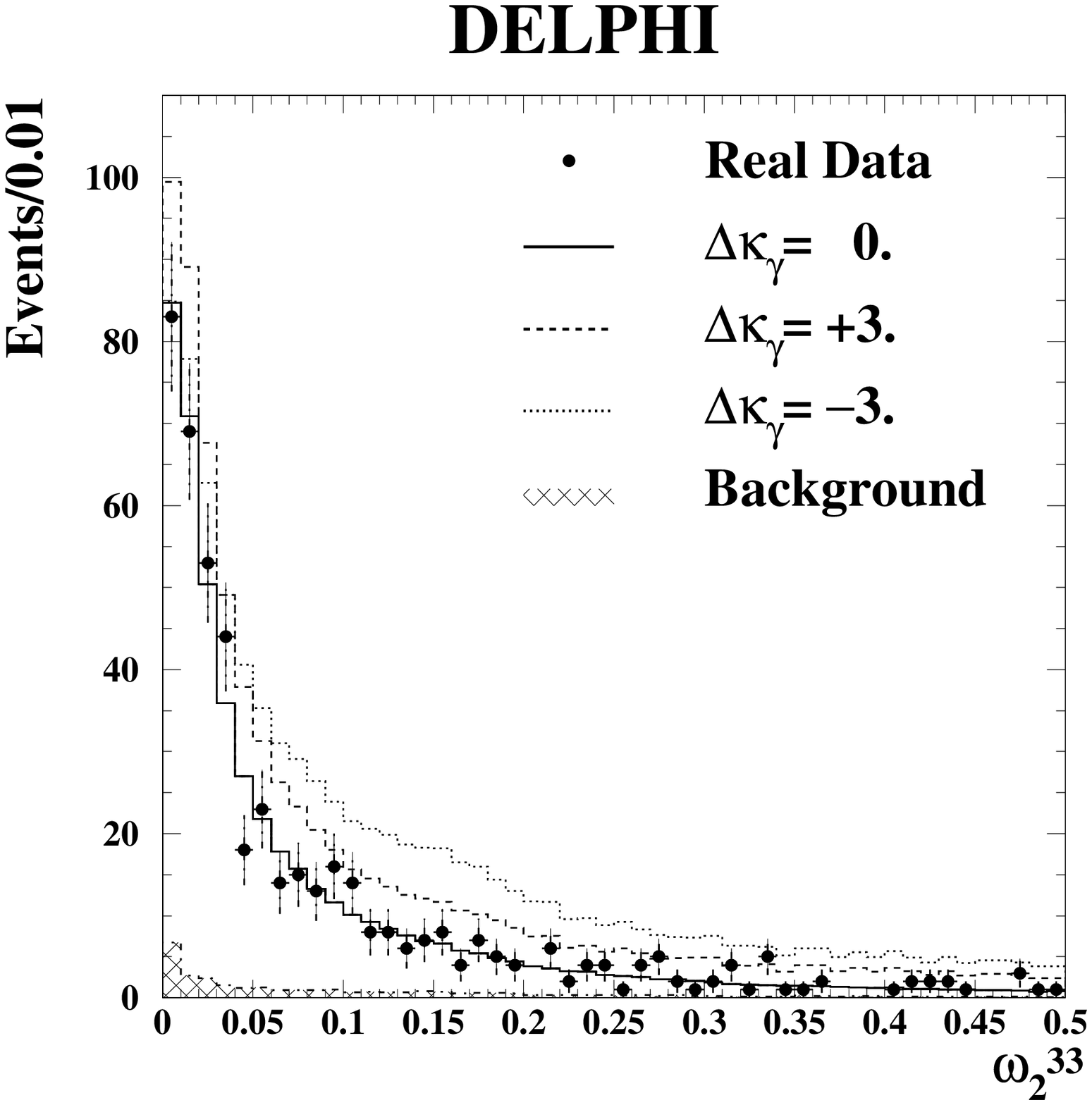,height=0.65\textwidth,bbllx=10,bblly=135,bburx=515,bbury=640,clip}
\caption
    {Distribution of the Optimal Variables $\omega_{1}^{3}$ and $\omega_{2}^{33}$ (the coefficients, respectively, of \dkg\ and  of (\dkg)$^2$ in the expansion of the differential cross-section in terms of Optimal Variables) for  semi-leptonic data at 200~GeV.  The points represent the real data, the solid lines the expected distributions for the SM value of the coupling, and the dashed lines the expected  distributions for the non-SM values \dkg = $\pm 3$.  The shaded area represents the background. The simulated distributions are normalised to the same luminosity as the data.}
\label{fig_OO_lg}
\end{figure}

\begin{figure}[!htbp]
\centering\epsfig{file=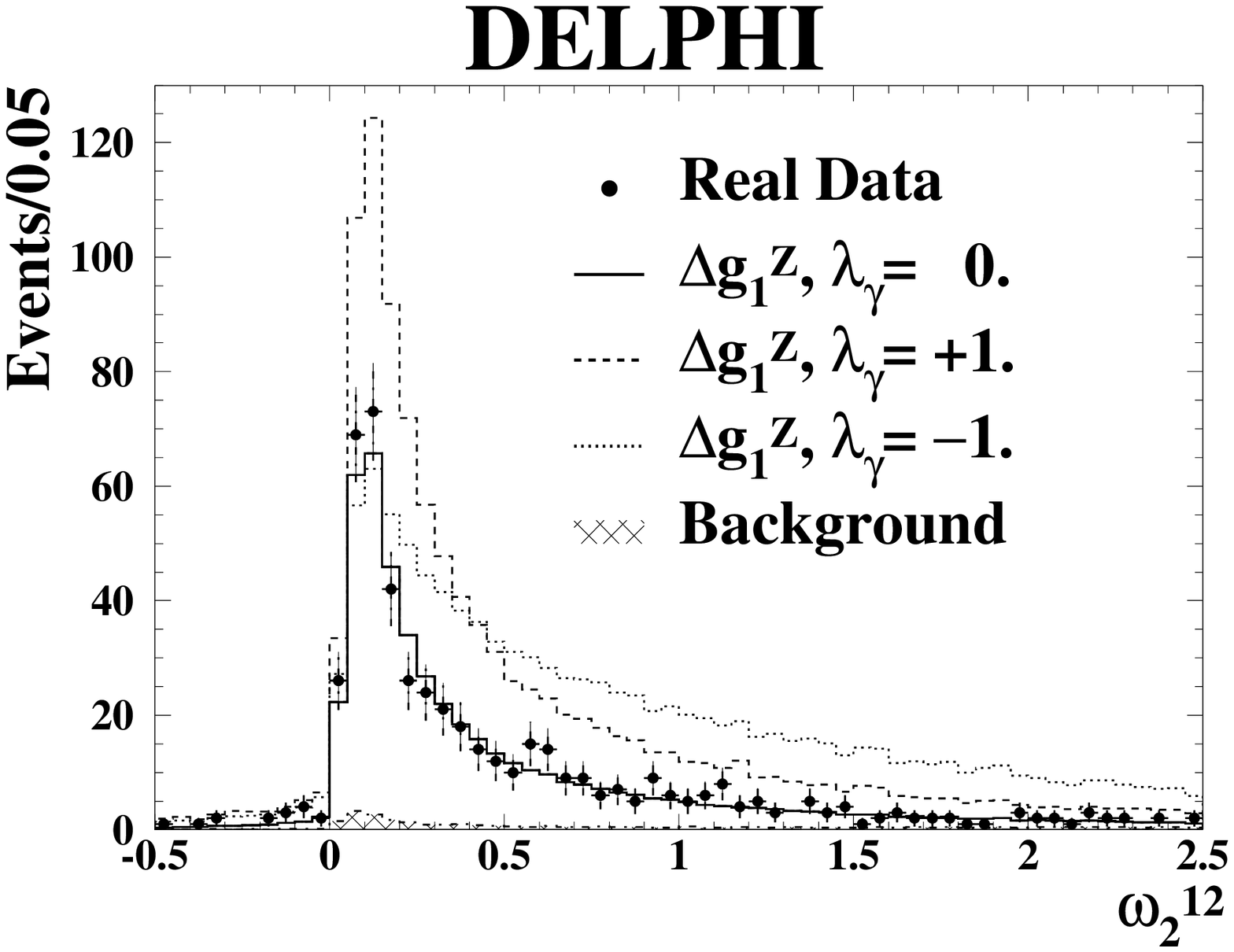,height=0.55\textwidth,bbllx=10,bblly=135,bburx=515,bbury=640,clip} \\
\vspace{-1.6cm}
\centering\epsfig{file=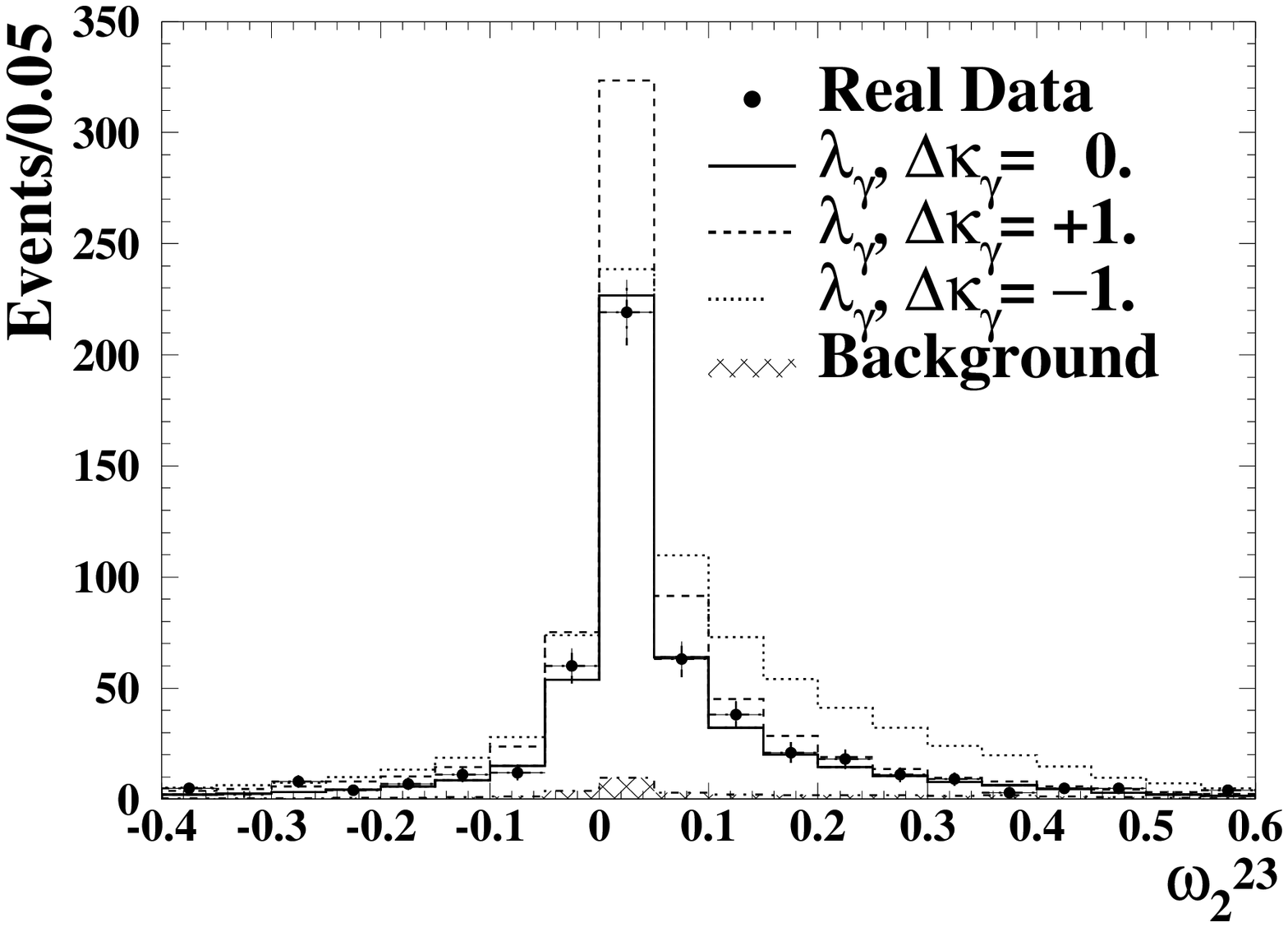,height=0.55\textwidth,bbllx=10,bblly=135,bburx=515,bbury=640,clip} \\
\vspace{-1.6cm}
\centering\epsfig{file=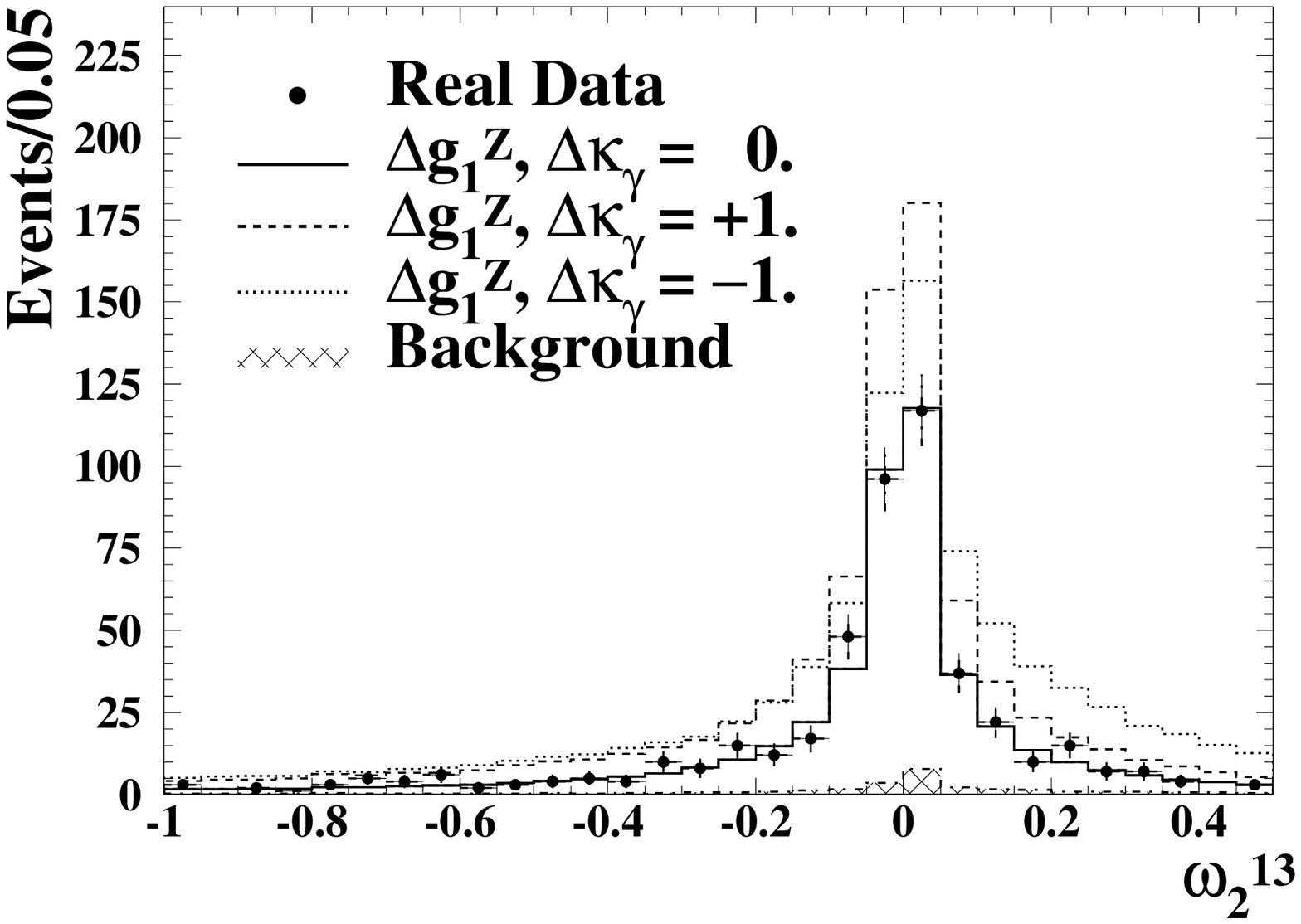,height=0.55\textwidth,bbllx=10,bblly=135,bburx=515,bbury=640,clip}
\vspace{-1.2cm}
\caption
    {Distribution of the Optimal Variables $\omega_{2}^{12}$, $\omega_{2}^{23}$ and $\omega_2^{13}$ (the coefficients, respectively, of \dgz$\cdot$\lgamma, \lgamma$\cdot$\dkg\ and \dgz$\cdot$\dkg\ in the expansion of the differential cross-section in terms of Optimal Variables) for semi-leptonic data at 200~GeV.  The points represent the real data, the solid lines the expected distributions for SM values of the couplings, and the dashed lines the expected  distributions for the non-SM values of the couplings shown in the legends. The shaded area represents the background. The simulated distributions are normalised to the same luminosity as the data.}
\label{fig_OO_xterm}
\end{figure}

The distributions of the Optimal Variables used in fits to the parameters \dgz\ ($i = 1$), \lgamma\ ($i = 2$) and \dkg\ ($i = 3$) are shown for the real data and for events simulated with SM and non-SM values of the couplings in figures~\ref{fig_OO_dkg} to \ref{fig_OO_xterm} for a centre-of-mass energy of 200~GeV.

The values of the coupling parameters were determined by binned extended maximum likelihood fits to the relevant Optimal Variables. A clustering technique was used to define the binning of the data, full details of which can be found in~\cite{BINNING}.  The method used the data points to divide the phase space into equiprobable, multidimensional bins. For each fit, a set of $d$ variables ($d$ = 2, 5 or 9, as described above) was required to describe an event completely and for $n$ events the clustering technique divides the $d$-dimensional space into $n$ bins, each centred on one data point. The available simulated events are then assigned to the bins by calculating the scalar distance $D_{kl}$ of each simulated event $k$ to each of the data points $l$,

\begin{equation}
D_{kl} = {(\vec{R_l}-\vec{r_k})}^T~M~{(\vec{R_l}-\vec{r_k})} \, ,
\label{eqn_Dkl}
\end{equation}

\noindent and assigning the $k$th simulated event to the bin $l$ for which $D_{kl}$ is a minimum. In~(\ref{eqn_Dkl}), $\vec{R}$ and $\vec{r}$ are the $d$-dimensional vectors that describe the real data point and the simulated event, respectively, and $M$ is a $d\times d$ matrix representing the metric of the space. The metric $M$ was defined by the variances and correlations of the real data distributions of the Optimal Variables being determined in any particular fit, so as to take into account the fact (observed in figures~\ref{fig_OO_dkg} to \ref{fig_OO_xterm}) that the different variables span different numerical ranges.

The technique described above assumes that the phase space variables $\Omega$ are fully determined for each event. In fact, one ambiguity remains for every event, namely that it is not known which of the jets from the hadronic $W$ decay comes from the quark, and which from the antiquark. In the analysis, each event was therefore entered twice into the maximum likelihood function, once with each of these two assignments.

\begin{figure}[!htbp]
\begin{center}
\epsfig{file=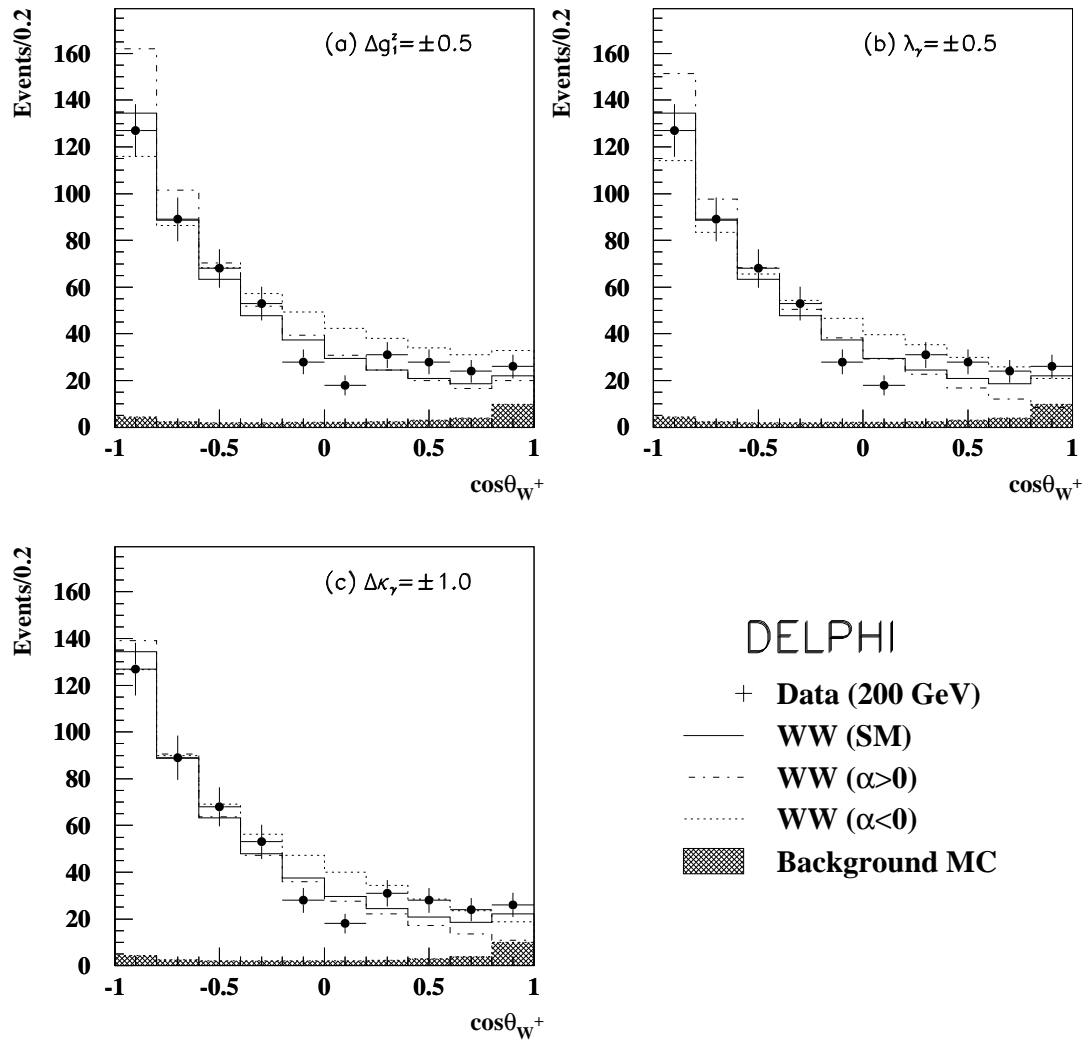,width=\textwidth} 
\end{center}
\caption
{The distribution of \ctw, the cosine of the polar angle of the \wplus\ in semi-leptonic events, at a centre-of-mass energy of 200~GeV. All three plots show the data (represented by points), the Standard Model prediction (the solid line) and the predicted background (the darker shaded region). Each plot also shows predictions for non-Standard Model values of a coupling $\alpha$: in a), $\alpha \equiv$~\dgz, in b) $\alpha \equiv$~\lgamma, and in c) $\alpha \equiv$~\dkg. The simulated distributions are normalised to the same luminosity as the data.} 
\label{fig_qqlv_ctw}
\end{figure}

\begin{figure}[!htbp]
\begin{center}
\epsfig{file=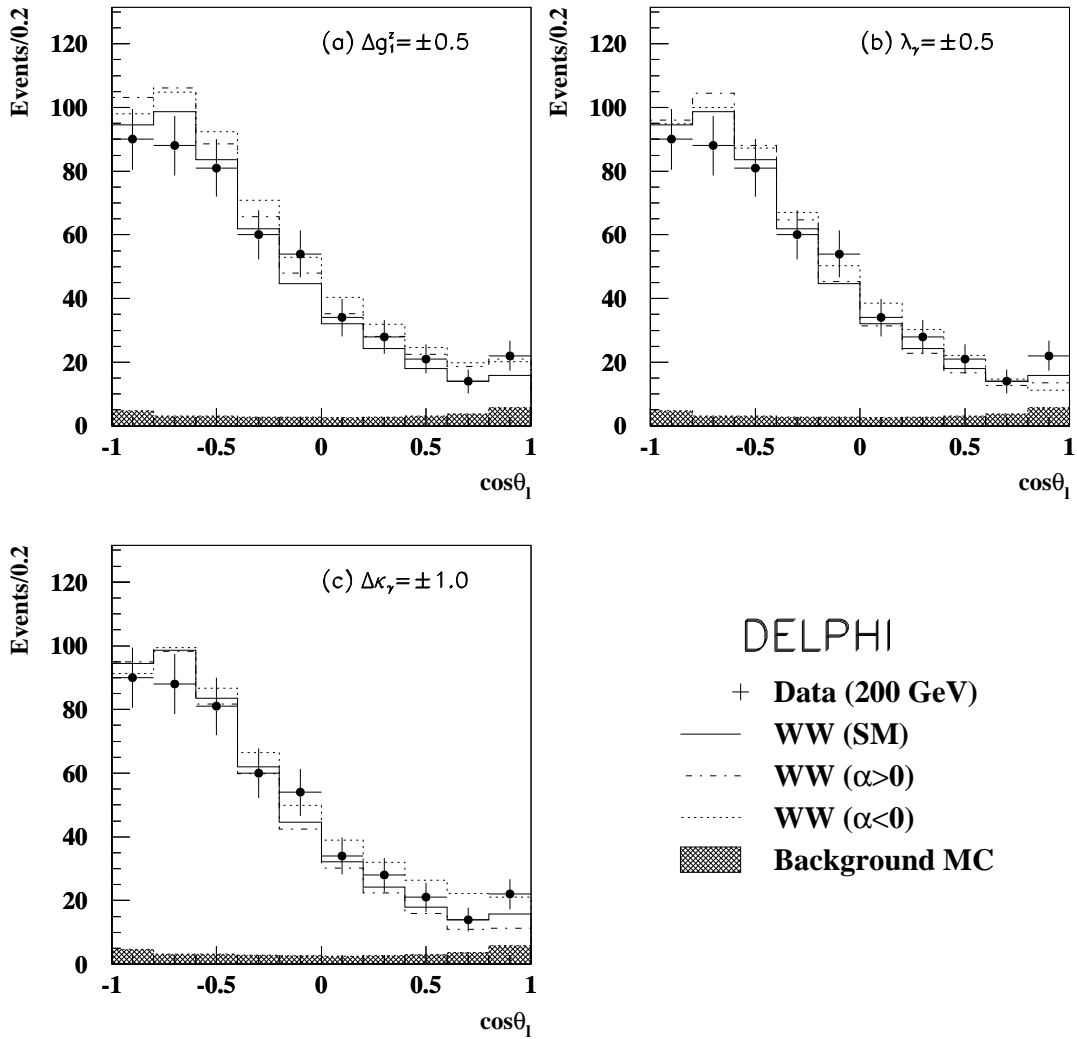,width=\textwidth} 
\end{center}
\caption
{The distribution of \ctl,  the cosine of the polar angle of the lepton in semi-leptonic events with respect to the incoming $e^\pm$ of the opposite sign, at a centre-of-mass energy of 200~GeV. All three plots show the data (represented by points), the Standard Model prediction (the solid line) and the predicted background (the darker shaded region). Each plot also shows predictions for non-Standard Model values of a coupling $\alpha$: in a), $\alpha \equiv$~\dgz, in b) $\alpha \equiv$~\lgamma, and in c) $\alpha \equiv$~\dkg. The simulated distributions are normalised to the same luminosity as the data.}
\label{fig_qqlv_ctl}
\end{figure}

\begin{figure}[!htbp]
\begin{center}
\epsfig{file=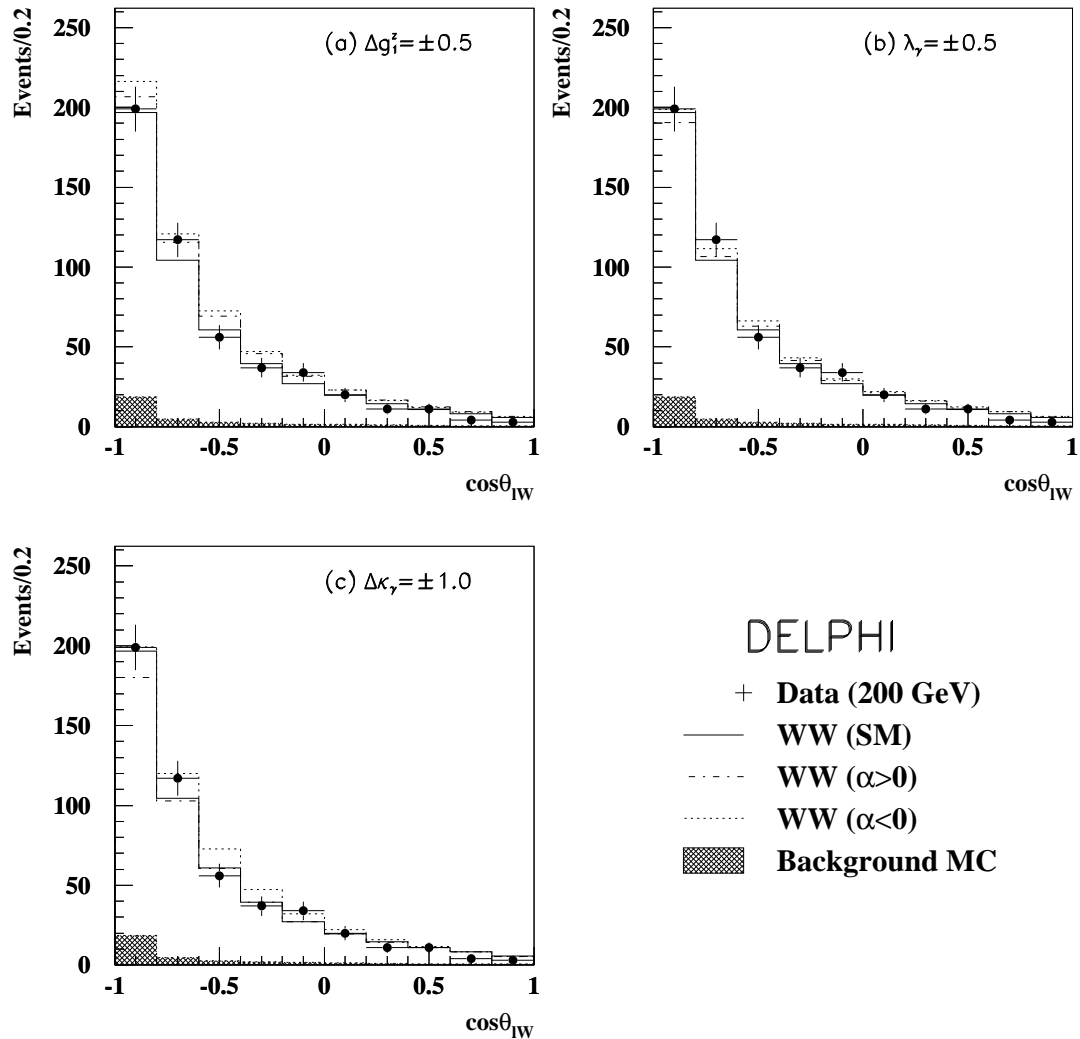,width=\textwidth} 
\end{center}
\caption
{The distribution of \clw, the cosine of the angle between the directions of the lepton and the hadronic $W$ in semi-leptonic events, at a centre-of-mass energy of 200~GeV. All three plots show the data (represented by points), the Standard Model prediction (the solid line) and the predicted background (the darker shaded region). Each plot also shows predictions for non-Standard Model values of a coupling $\alpha$: in a), $\alpha \equiv$~\dgz, in b) $\alpha \equiv$~\lgamma, and in c) $\alpha \equiv$~\dkg. The simulated distributions are normalised to the same luminosity as the data.}
\label{fig_qqlv_clw}
\end{figure}

A second analysis was performed in the \jjlv\ channel as a cross-check. In this analysis a binned maximum likelihood fit was made to the differential cross-section of three angles: \ctw, the $W^+$ production angle, \ctl, the polar angle of the lepton with respect to the incoming $e^\pm$ of the opposite sign, and \clw, the cosine of the angle between the hadronic W and the lepton.  The same event selection criteria were applied and the same re-weighting method  was used as in the Optimal Variable analysis. The distributions of these angular variables are shown in figures~\ref{fig_qqlv_ctw} to~\ref{fig_qqlv_clw} for the data and for events simulated with different values of the couplings at 200~GeV.

\subsection{Fully hadronic final state}
\label{sec_anal_jjjj}

The analysis of events in the fully hadronic state is complicated by the fact that the four observed hadronic jets cannot immediately be assigned to a particular $W^+$ or $W^-$ decay. Two problems arise from this feature, first, that it is not clear which of the three possible pairings of the four jets corresponds to a $WW$ pair, and, second, once the pairing is decided, which of the di-jet pairs is the $W^+$ and which the $W^-$. 

The first of these problems was approached by forcing the selected events into a four-jet configuration and  constructing a neural network  to determine the combination which was most likely to represent a $W$ pair event.  A kinematic fit, imposing four-momentum conservation and equal mass for the two di-jet pairs, was performed for each of the three combinations. The $\chi^2$ of the kinematic fit and the difference between the nominal $W$ mass and the di-jet mass from the fit were used as inputs to the neural network to choose the most likely combination. The efficiency of this procedure was estimated to be about~79\%, where the uncertainty in the pairing was estimated by repeating the procedure using simulated events generated with the different parton shower and fragmentation models implemented in PYTHIA, HERWIG~\cite{HERWIG} and ARIADNE~\cite{ARIADNE}.

The second problem -- to distinguish which of the di-jet pairs came from the $W^+$ and $W^-$ -- was partly resolved by
constructing an effective jet charge $Q_{jet}$ from the charge of the particles in the jet, weighted by their momentum:

\begin{equation}
Q_{jet}=\frac{\sum_i q_i (\vec{p}_i\cdot\vec{T}_{jet})^{0.7}} {\sum_i (\vec{p}_i\cdot\vec{T}_{jet})^{0.7}} \, ,
\end{equation}  

\noindent where $q_i$ and $p_i$ are, respectively, the charge and momentum of the particle in the jet, $\vec{T}_{jet}$ 
is the unit vector in the reconstructed jet direction and the exponent 0.7 was chosen empirically.  Then, following the method described in~\cite{ALEPHjjjj}, the charge difference of the two di-jet pairs, 

\begin{equation}
\Delta Q = (Q_{jet_1}+Q_{jet_2})-(Q_{jet_3}+Q_{jet_4}) \, ,
\end{equation}

\noindent was used to assign the charge of the individual $W^{\pm}$ bosons.  The more negative di-jet  was tagged as originating from a $W^-$, and the other di-jet as the $W^+$. The efficiency of this procedure was estimated from the simulation to be about~76\% for events with correct jet pairing, using the minimal angle between the reconstructed di-jet system and the generated $W$ boson to determine the correct pairing. As in the case of the jet pairing studies described above, the systematic uncertainty of this procedure was estimated by using the different parton shower and fragmentation models implemented in PYTHIA, HERWIG and ARIADNE.

The ambiguity in the charge was taken into account by constructing a new variable:
 
\begin{equation}
x_q = P_{W^-}(\Delta Q)\cos\theta_{W^-} - (1-P_{W^-}(\Delta Q))\cos\theta_{W^-} \, ,
\end{equation}

\noindent where $\cos\theta_{W^-}$ is the polar angle of the di-jet pair assigned to the $W^-$ and $P_{W^-}(\Delta Q)$ is the probability that the di-jet pair originates from a $W^-$.  The value of $P_{W^-}(\Delta Q)$ was obtained from the distribution of $\Delta Q$ in the simulated events. The couplings were then estimated from a binned extended maximum likelihood fit to the variable $x_q$.  

\subsection{Single W final state}

In the \jjX~final state, the couplings were extracted via a binned maximum likelihood fit to the distribution of the angle between the jets. This is 
a  well-measured variable, and was found to be more sensitive to the coupling parameters than, say, the $W$ production angle (a result which follows from the dynamics of the Feynman diagram (figure~\ref{fig_tgc}c) providing the dominant contribution to the sensitivity to the couplings in the \jjX\ sample). The \lX~final state was analysed using a maximum likelihood fit to the number of events selected in the data, no further sub-division of the data being found to give a significant improvement to the experimental sensitivity. As mentioned in sections~\ref{sec_sel_jjX} and~\ref{sec_sel_lX}, the samples selected in these final states include contributions from some processes labelled as ``background", but nonetheless with trilinear gauge couplings involved in their production mechanisms; in the fits performed, the relevant parameters were varied wherever they occurred in the production processes contributing to the events expected in the selected samples. 

Since only the $WW\gamma$ vertex occurs in the production of the \Wev\ final state via a trilinear gauge coupling (as seen in  figure~\ref{fig_tgc}(c)), the sensitivity of the single~$W$ channels to \dgz\ is very poor, and fits to this parameter were not used in the results presented. The likelihood distributions from fits of the other two parameters, \lgamma\ and \dkg, to the \jjX\ and \lX\ final states were combined, and the resulting distributions were subsequently combined with those from the \jjlv\ and \jjjj\ final states in the determination of the coupling parameters.

\section{Systematic uncertainties}
\label{sec_systematics}

Sources of systematic uncertainty were considered which contribute to the results in all the final states analysed. Those arising in the analysis of the final states from $WW$ production are described in section~\ref{sec_systematics_WW}; the contribution to the total uncertainty from each source to the results for each of the three coupling parameters determined from data in the \jjlv\ and \jjjj\ channels is given in tables~\ref{tab_systematics_jjlv} and~\ref{tab_systematics_jjjj}, respectively. A similar study was performed for the couplings \lgamma\ and \dkg\ determined from data in the single $W$ final states. A summary is given in section~\ref{sec_systematics_singleW}  and the results are reported in table~\ref{tab_systematics_singleW}.

\subsection{WW final states} 
\label{sec_systematics_WW}

\subsection*{W pair production cross-section and radiative corrections:}

The calculation of the $W$ pair production cross-section was significantly improved in 2000~\cite{LEP2MCWorkshop}.  The theoretical uncertainty in the relevant energy range was reduced from the level of 2\%~\cite{Gentle2} quoted in previous publications~\cite{DELPHI_TGC189} to 0.5\%~\cite{YFSWW,RacoonWW} via the inclusion of all ${\cal O}(\alpha)$ corrections. The systematic uncertainties in the coupling parameters arising from this latest estimate of the uncertainty in the total $WW$ cross-section are shown in the tables.  

The inclusion of the ${\cal O}(\alpha)$ corrections has also been shown to have a marked effect on the differential distributions~\cite{RAD_CORR}, which could substantially affect the measurement of the gauge boson coupling parameters. The determination of the resulting systematic uncertainty in the determination of the couplings required the use of re-weighted events. The weights were generated using YFSWW in the simulation and were used according to the procedure described in~\cite{WPHAC_SETUP}.  The effect on the measurement of the couplings arising from the theoretical uncertainty in the calculation of the radiative corrections was then obtained in two stages. First, one million fully simulated Monte Carlo events were produced at 189~GeV using the generators WPHACT, RacoonWW~\cite{RacoonWW} and YFSWW.  From a comparison of the couplings determined from analysis of these different samples, it was possible to estimate the systematic uncertainty from higher order electroweak corrections on the calculation of the Initial State Radiation.  This was found to be negligible. Then, using the sample simulated with YFSWW, a comparison was made of two different Leading Pole approximation schemes, the so-called LPA-A and LPA-B schemes. The differences in the couplings determined from analysis of samples employing these two models was taken to represent the systematic error from the uncertainty of the dependence of the Double Pole Approximation on the assumed LPA scheme.  It can be seen in the corresponding entries in tables~\ref{tab_systematics_jjlv} and~\ref{tab_systematics_jjjj} that this uncertainty gives rise to significant systematic errors in the measurement of the couplings.

\begin{table}[!htbp]
\begin{center}
\begin{tabular}{llll}                                                  
\hline
\multicolumn{1}{c}{Source} & \multicolumn{1}{c}{\dgz}  & \multicolumn{1}{c}{\lgamma} & \multicolumn{1}{c}{\dkg}  \\ 
\hline
\hline
$WW$ cross-section           & 0.0005 & 0.0006 & 0.007  \\
Radiative Corrections          & $_{-0.002}^{+0.005}$ & $_{-0.002}^{+0.004}$ & $_{-0.015}^{+0.012}$ \\
Background cross-section   & 0.004  & 0.003  & 0.014  \\
$W$ Mass                            & 0.001  & 0.001  & 0.002   \\
LEP beam energy                & 0.0005 & 0.0005 & 0.001 \\    
Luminosity                            & 0.0005 & 0.0006 & 0.007  \\
Fragmentation                      & 0.005  & 0.005  & 0.015  \\
\hline
Lepton tagging efficiency     & 0.003  & 0.003  & 0.001  \\
Lepton charge assignment   & 0.005  & 0.005  & 0.003  \\
Jet reconstruction                & 0.002  & 0.002  & 0.007  \\
Lepton reconstruction         & 0.001  & 0.001  & 0.003  \\
\hline 
Total                                    & $_{-0.009}^{+0.010}$ & $_{-0.009}^{+0.010}$ & $_{-0.028}^{+0.027}$ \\
\hline
\hline 
Statistical errors                 & $_{-0.031}^{+0.033}$ & $_{-0.035}^{+0.036}$ & $_{-0.094}^{+0.103}$ \\
\hline
\end{tabular}
\caption[]
{Contributions to the systematic errors on the couplings determined from data in the semi-leptonic final state, \jjlv. Except where otherwise indicated, the errors are symmetric with respect to a change of sign of the parameters involved. The first 7 sources listed in the table are considered to be fully correlated with the other channels. For comparison, the bottom row of the table lists the statistical errors on the couplings determined in the Optimal Variables analysis (also shown in table~\ref{tab_results_1d}). }
\label{tab_systematics_jjlv}
\end{center}
\end{table}
\begin{table}[!htbp]
\begin{center}
\begin{tabular}{lccc}                                                  
\hline
\multicolumn{1}{c}{Source} & \dgz  & \lgamma & \dkg  \\ 
\hline
\hline
$WW$ cross-section         & 0.006 & 0.008 & 0.011 \\
Radiative corrections    & 0.017 & 0.016 & 0.032 \\
Background cross-section & 0.003 & 0.004 & 0.009 \\
$W$ mass                   & 0.003 & 0.003 & 0.005 \\
LEP beam energy          & 0.001 & 0.001 & 0.001 \\
Luminosity               & 0.006 & 0.008 & 0.011 \\
Fragmentation            & 0.009 & 0.012 & 0.027 \\
\hline                                      
Colour Reconnection      & 0.008 & 0.006 & 0.012 \\
Bose Einstein            & 0.002 & 0.002 & 0.005 \\
Simulation statistics            & 0.008 & 0.009 & 0.012 \\
Selection efficiency     & 0.005 & 0.005 & 0.007 \\
Event reconstruction     & 0.004 & 0.004 & 0.008 \\
\hline                                      
Total                                   & 0.024                          & 0.025                         & 0.049 \\
\hline
\hline
Statistical errors                 & $_{-0.067}^{+0.083}$ & $_{-0.070}^{+0.093}$ & $_{-0.149}^{+0.196}$ \\ 
\hline
\end{tabular}
\caption[]
{Contributions to the systematic errors on the couplings determined from data in the fully hadronic final state, \jjjj. The first 7 sources listed in the table are considered to be fully correlated with the other channels. For comparison, the bottom row of the table lists the statistical errors on the couplings (also shown in table~\ref{tab_results_1d}).}
\label{tab_systematics_jjjj}
\end{center}
\end{table}
\begin{table}[!htbp]
\begin{center}
\begin{tabular}{lll}                                                  
\hline
\multicolumn{1}{c}{Source} & \multicolumn{1}{c}{\lgamma} & \multicolumn{1}{c}{\dkg}  \\ 
\hline
\hline
Signal cross-section            & 0.005 & 0.037   \\
Background cross-section  &  0.002 & 0.002  \\
Selection efficiency              & 0.011  & 0.072   \\
\hline
Total                                      & 0.011                      & 0.081  \\
\hline
\hline 
Statistical errors                  & $_{-0.288}^{+0.250}$ & $_{-0.148}^{+0.131}$ \\
\hline
\end{tabular}
\caption[]
{Contributions to the systematic errors on the couplings determined from data in the single $W$ final states. For comparison, the bottom row of the table lists the statistical errors on the couplings (also shown in table~\ref{tab_results_1d}).}
\label{tab_systematics_singleW}
\end{center}
\end{table}

\subsection*{Background cross-sections and modelling:}

The theoretical uncertainty on the cross-sections of two- and four-fermion 
processes varies between 2\% and 5\%, depending on the process. A
conservative estimate of the systematic error on the couplings was made
by varying the predicted background cross-sections by $\pm$5\%.

\subsection*{W mass and LEP beam energy:}

The systematic error arising from the uncertainty on the $W$ mass used in the event simulation was evaluated using data samples generated with masses 1~GeV/$c^2$ above and below the nominal value.  A linear interpolation was used to scale the systematic error to that which would arise from an uncertainty in the $W$ mass of $\pm$40~MeV/$c^2$.

The same method was used to estimate the systematic uncertainty due to the value of the \LEP~beam energy used in the simulation; samples were generated with different centre-of-mass energies and the errors were rescaled to the measured beam energy uncertainties~\cite{LEP_BEAM}.

\subsection*{Determination of the luminosity:}

The luminosity was determined from a measurement of Bhabha
scattering and was affected by the experimental error on the
acceptance ($\pm$0.5\%) and the theoretical uncertainty on the cross-section
($\pm$0.12\%)~\cite{BHABHA}.  The estimated uncertainty on the luminosity
was used to vary the normalisation of the simulation in the fits.

\subsection*{Modelling of fragmentation:}

In order to assess the effect of the model used for the fragmentation of hadronic jets -- JETSET final state QCD radiation and fragmentation, -- correlated samples were analysed using the modelling of HERWIG and ARIADNE, and the differences in the fitted values of the coupling parameters noted. The largest discrepancies found were between JETSET and HERWIG and these were taken as a conservative estimate in each channel.

Additional tests were performed in the fully hadronic final state using mixed Lorentz-boosted $Z$ events~\cite{MLBZ}, in which $WW$ events are emulated using two events taken at the $Z$~peak, and transforming them such that their superposition reflects that of a true fully hadronic $WW$ event. These studies are also sensitive to systematic errors in the event reconstruction technique, and are discussed further in the relevant section below.

\subsection*{Final state interactions:}

The measurement of the couplings in the fully hadronic final state is affected by final state interactions between the decay products of the two $W$ bosons.  Two effects were considered: the exchange of gluons between the quarks of different $W$ bosons, known as Colour Reconnection, and Bose-Einstein correlations between pions.

\subsubsection*{Colour Reconnection:}

In the reaction $e^+ e^- \rightarrow W^+W^- \rightarrow (q_1\bar{q}_2)(q_3\bar{q}_4)$ the hadronisation models used in this analysis treat the colour singlets $q_1\bar{q}_2$ and $q_3\bar{q}_4$ coming from each $W$ boson independently. However, interconnection effects between the products of the two $W$ bosons may be expected since the mean $W$ lifetime is an order of magnitude smaller than the typical hadronisation times. This can lead to the exchange of coloured gluons between partons from the hadronic systems from different $W$ bosons - the Colour Reconnection effect - in the development of the parton showers. This, in turn, can give rise to a distortion in the angular distributions of the final hadronic systems used to estimate the primary quark directions in the determination of the triple gauge coupling parameters from \jjjj\ data. These effects can be large at hadronisation level, due to the large numbers of soft gluons sharing the space-time region, and have been studied by introducing colour reconnection effects into various hadronisation models. The most studied model is the Sj\"{o}strand-Khoze ``Type 1'' model (SKI)~\cite{SK1}, and this was used for the evaluation of the systematic uncertainty in the analysis reported here. The model is based on the Lund string fragmentation phenomenology, in which the volume of overlap between two strings, and hence the colour reconnection probability, is represented by a parameter, $\kappa$.

In this paper, the systematic uncertainty was estimated using the SKI model with full colour reconnection ({\it i.e.} $\kappa = \infty$). This is a highly conservative assumption when compared with the direct measurements of colour reconnection reported by \DELPHI~\cite{CR} and by other LEP experiments~\cite{L3CR,OPALCR,ALEPHCR}.  Symmetric systematic errors were applied to the gauge coupling parameters, representing the difference observed between full colour reconnection and no effect from this source.

\subsubsection*{Bose-Einstein correlations:}

Correlations between final state hadronic particles are dominated by Bose-Einstein correlations, a quantum mechanical effect which enhances the production of identical bosons close in phase space. The net effect is that multiplets of identical bosons are produced with smaller energy-momentum differences than non-identical ones. This, again, can affect the estimation of the primary quark directions in data from hadronically decaying $W$ bosons. Bose-Einstein correlations between particles produced from the same $W$ boson affect the normal fragmentation and are treated implicitly in the fragmentation uncertainties constrained by data from $Z$ decays, while correlations between pairs of particles coming from different $W$ bosons cannot be constrained or safely predicted by the information from single hadronically decaying vector bosons, and are estimated in various models. We have used the LUBOEI~BE$_{32}$ model~\cite{LUBOEI} to estimate the systematic uncertainty in the determination of gauge coupling parameters from the present data. In this model, Bose-Einstein correlations are described using two parameters: the correlation strength, $\lambda$, and the effective source radius, $R$. Applying the model with parameters $\lambda=1.35$ and $R=0.6$~fm, symmetric systematic errors on the gauge coupling parameters were estimated by taking the difference between the values obtained assuming the presence of Bose-Einstein correlations only within each $W$ and those obtained assuming correlations both within and between $W$ bosons. Taking into account the reported results of measurements of Bose-Einstein correlations by \DELPHI~\cite{DBEC} and in other LEP~\cite{L3BEC,OPALBEC,ALEPHBEC} experiments, this again represents a conservative estimate of the effect from this source.

\subsection*{Statistics of simulated samples and selection efficiency:}

The statistical error on the number of simulated events assigned to each data bin was convoluted in the fitting method for fits to the data in the semi-leptonic channel; the fitting method ensures that this systematic error is negligible with the large statistics available.  In the fully hadronic channel, the distribution of simulated events used in the binned extended maximum likelihood fit was varied according to the statistical uncertainties of the bin contents.

The uncertainty due to the event selection efficiency was used to vary the normalisation of the simulation in the fits.
 
\subsection*{Lepton tagging efficiency and charge assignment:}

Comparisons were made between fully simulated events and real $Z$ events to estimate the possibility of having different lepton tagging efficiencies in the data.  The systematic uncertainty was estimated assuming 1\% mis-tagging for muons and for electrons in the barrel region and 5\% for electrons in the forward region of the detector. The value shown in table~\ref{tab_systematics_jjlv} represents the combined effect from both lepton types, with the dominant contribution coming from mis-tagged electrons. However, the effect is reduced as mis-tagged electrons or muons can be retrieved by the single prong tau selection. 

The effect of wrongly assigned lepton charge was estimated using data simulated at the $Z$ pole by counting the numbers of di-lepton events in which the two leptons are assigned the same charge.  A mis-assignment rate of 0.1\% was found for all lepton candidates except for electrons in the forward region, where the rate rose to 6\%. The systematic error was calculated by randomly changing the charge of the lepton candidate in the fits with these probabilities, and the value shown in table~\ref{tab_systematics_jjlv} shows the combined effect of these assumed uncertainties.

\subsection*{Event reconstruction:}

The effect of possible systematic errors in the event reconstruction technique was estimated using comparisons between data and simulation. This was performed in two ways: firstly, by comparing significant variables used in the analysis in data and simulation and computing the effect of the discrepancy seen; and secondly, by directly computing changes in the results using mixed Lorentz-boosted $Z$ (MLBZ) events, mentioned above in the section on systematic errors resulting from the modelling of fragmentation.  

In the semi-leptonic channel,  the systematic uncertainty in the couplings due to uncertainties in the lepton and jet energies and angular distributions was estimated using comparisons between data and simulated events at the $Z$ peak.  The estimated uncertainties on the jet energies and angles were found to be 5\% and 7.5 mrad, respectively.  The uncertainty on the muon momentum was found to be 1\%, while for electron momenta uncertainties of 1\% and 5\%  were estimated in the barrel and forward regions, respectively.  Appropriate smearings were applied to these resolutions in the simulation of \jjlv\ events and the resulting shifts in the values of the couplings were taken to be the systematic uncertainties. They are reported in table~\ref{tab_systematics_jjlv}  as the systematic errors arising from jet and lepton reconstruction.

In the fully-hadronic channel, the uncertainties in the event reconstruction were estimated using MLBZ events from both real and simulated data at the $Z$ peak. As described above in the discussion of the modelling of quark fragmentation, the MLBZ method emulates $WW$ events using two events taken at the $Z$~peak, rotating them and Lorentz-boosting them such that their superposition reflects that of a true $WW$ event. The detector effects are thus included in as realistic a manner as possible. In order to estimate these effects on the determination of gauge coupling parameters in $WW \ra$~\jjjj\ events, the ratio of selection efficiencies, $r$, of MLBZ data events to MLBZ simulated events was determined as a function of the simulated $W$ production angle. The ratio $r(\cos \theta_{W^-})$ was then applied to simulated $WW$ samples and the gauge coupling analysis described in section~\ref{sec_anal_jjjj}, which uses the $W$ production angle, was repeated. The differences between the results with and without application of the ratio were taken as systematic errors and are reported in table~\ref{tab_systematics_jjjj}. The systematic uncertainty evaluated by this method represents a conservative estimate, as it includes both the inaccuracies in the modelling of detector effects and most of the deviations induced by the applied fragmentation model.

An additional problem, not included in the effects considered above, has been encountered in the reconstruction of charged tracks in the forward region of \DELPHI~\cite{MLBZ}, leading to a small error in the reconstructed direction of forward tracks in both simulated and real data. Its effects were shown to be negligible in a previous \DELPHI\ analysis~\cite{DELPHI_SDM} involving fits to binned data of production and decay distributions in $WW$ production, and, in a study of the current data in the \jjlv\ final state at 200~GeV, have also been found to be negligible in comparison to the other correlated systematic errors considered. No systematic error has therefore been included from this source. 

\subsection{Single W final states}
\label{sec_systematics_singleW}

Systematic errors arising from the uncertainty in the signal (\Wev) cross-section were estimated by varying the cross-section by $\pm$5\% and noting the effect on the fitted coupling parameters. Similarly, cross-sections of other contributing channels were varied by $\pm$2\%, and the fits repeated. The maximum changes to the fitted parameters in the combined \jjX\ and \lX\ final states were taken as systematic errors, and are reported in table~\ref{tab_systematics_singleW} as the contributions from signal and background cross-sections, respectively. Systematic errors arising from the uncertainty in the selection efficiency were estimated from the statistical errors in the numbers of simulated events, and are also reported in the table. No other significant source of systematic error was identified in these channels. 

\section{Results}
\label{sec_results}

The procedure used to combine the results from the three channels and the results obtained are described in the following sections.

\subsection{Combination procedure}
\label{sec:gc_combination}

The combination was based on the individual likelihood functions from the samples of the three final states, \jjlv, \jjjj\ and \Wev, included in the analysis. Each final state sample provides the negative log likelihood, -$\ln {\mathcal L}$, at each centre-of-mass energy, as a function of the coupling parameters for inclusion in the combination.  

The $\ln {\mathcal L}$ functions from each channel include statistical errors as well as those systematic uncertainties which are considered as uncorrelated between channels.  For both single- and multi-parameter combinations, the individual $\ln {\mathcal L}$ functions were combined.  It is necessary to use the $\ln {\mathcal L}$ functions directly in the combination, since in some cases they are not parabolic, as discussed extensively in~\cite{BORUT}, and hence it is not possible to combine the results properly by simply taking weighted averages of the measurements.

The following sources of systematic uncertainty were  assumed to be correlated between the semi-leptonic and fully hadronic channels: $WW$ cross-section, radiative corrections, background cross-section, $W$ mass, beam energy, luminosity and fragmentation. The procedure used was based on the introduction of an additional free parameter to take into account each correlated source of systematic uncertainty. These parameters are treated as shifts on the fitted parameter values, and are assumed to have Gaussian distributions. A simultaneous minimisation of both sets of parameters (coupling parameters and systematic uncertainties) was performed on the log-likelihood function. 

In detail, the combination proceeded in the following way: the set of measurements from the three channels \jjlv, \jjjj\ and single $W$ is given with statistical plus uncorrelated systematic uncertainties in terms of likelihood curves  
$-\ln{\mathcal L}^{qql\nu}_{stat}(x)$, $-\ln{\mathcal L}^{qqqq}_{stat}(x)$ and $-\ln{\mathcal L}^{single W}_{stat}(x)$, respectively, where $x$ is the coupling parameter in question. Also given are the shifts for each of the totally correlated sources of uncertainty mentioned above, each source $S$ giving rise to systematic errors $\sigma^S_{qql\nu}$ and $\sigma^S_{qqqq}$. Additional parameters $\Delta^S$ are then included in the likelihood sum in order to take into account a Gaussian distribution for each of the systematic uncertainties. The procedure then consisted in minimising the function

\begin{eqnarray}
-\ln {\mathcal L}_{total} = -\sum_{C} \ln {\mathcal L}^C_{stat} (x-\sum_{S}(\sigma^S_C \Delta^S))
 + \sum_{S} {\displaystyle \frac{(\Delta^S)^{2}}{2}} \, ,
\end{eqnarray}

\noindent where $x$ and the $\Delta^S$ are the free parameters, the sum over $C$ runs over the three channels and the sum over $S$ runs over the seven sources of correlated systematic uncertainty. The resulting uncertainty on $x$ takes into account all sources of uncertainty, yielding a measurement of the coupling with a precision which includes the errors from both statistical and systematic sources. The projection of the minima of the log-likelihood as a function of $x$ gives the combined log-likelihood curve including statistical and systematic uncertainties. 

\subsection{Results}

\begin{figure}[!htbp]
\centering\epsfig{file=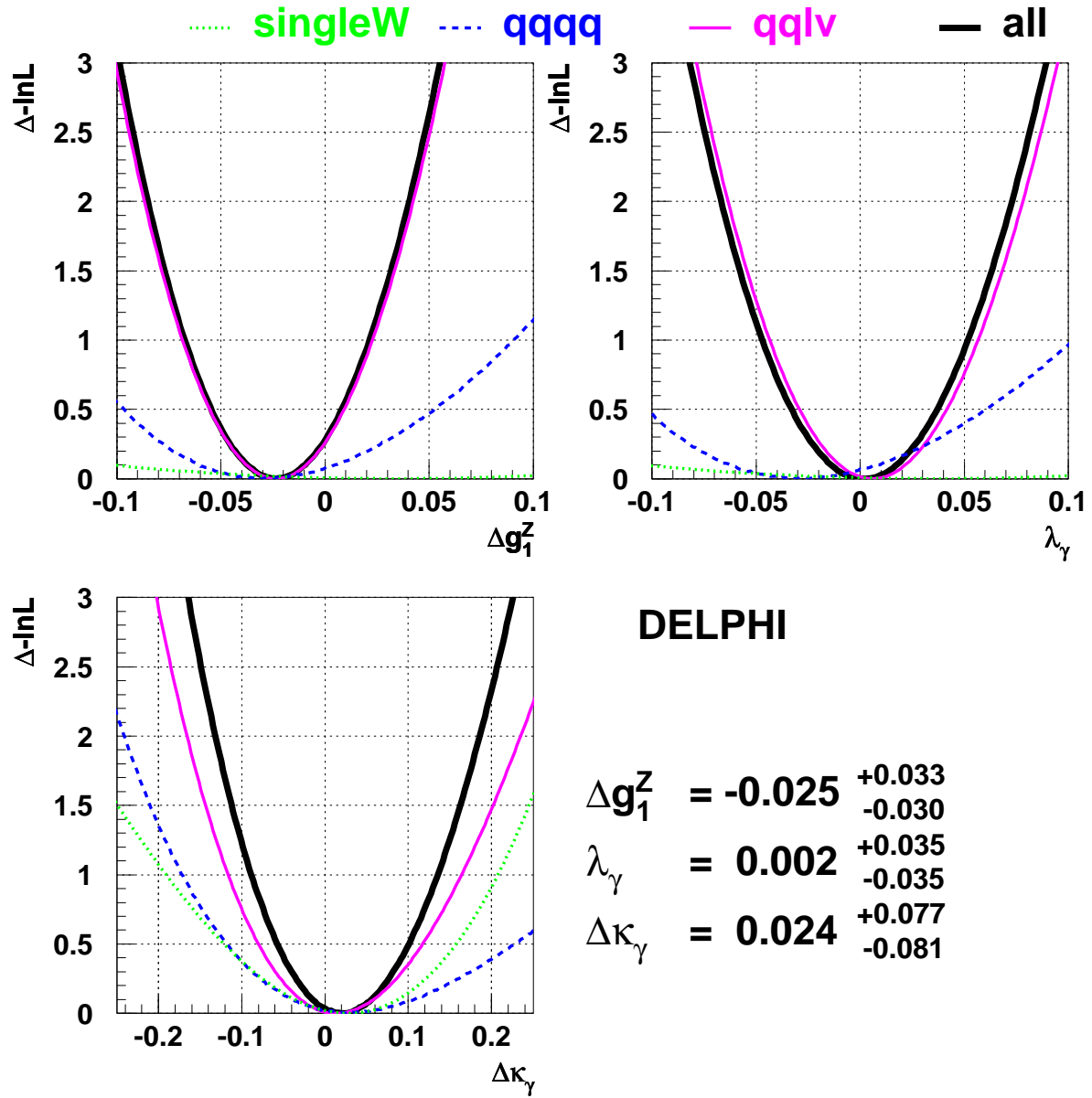,height=\textwidth}
\caption
    {The log-likelihood curves from single parameter fits to the data, combining results from the semi-leptonic, fully hadronic and single $W$ final states.The curves include contributions from both statistical and systematic effects}
\label{fig_results_1d}
\end{figure}

\begin{figure}[!htbp]
\centering\epsfig{file=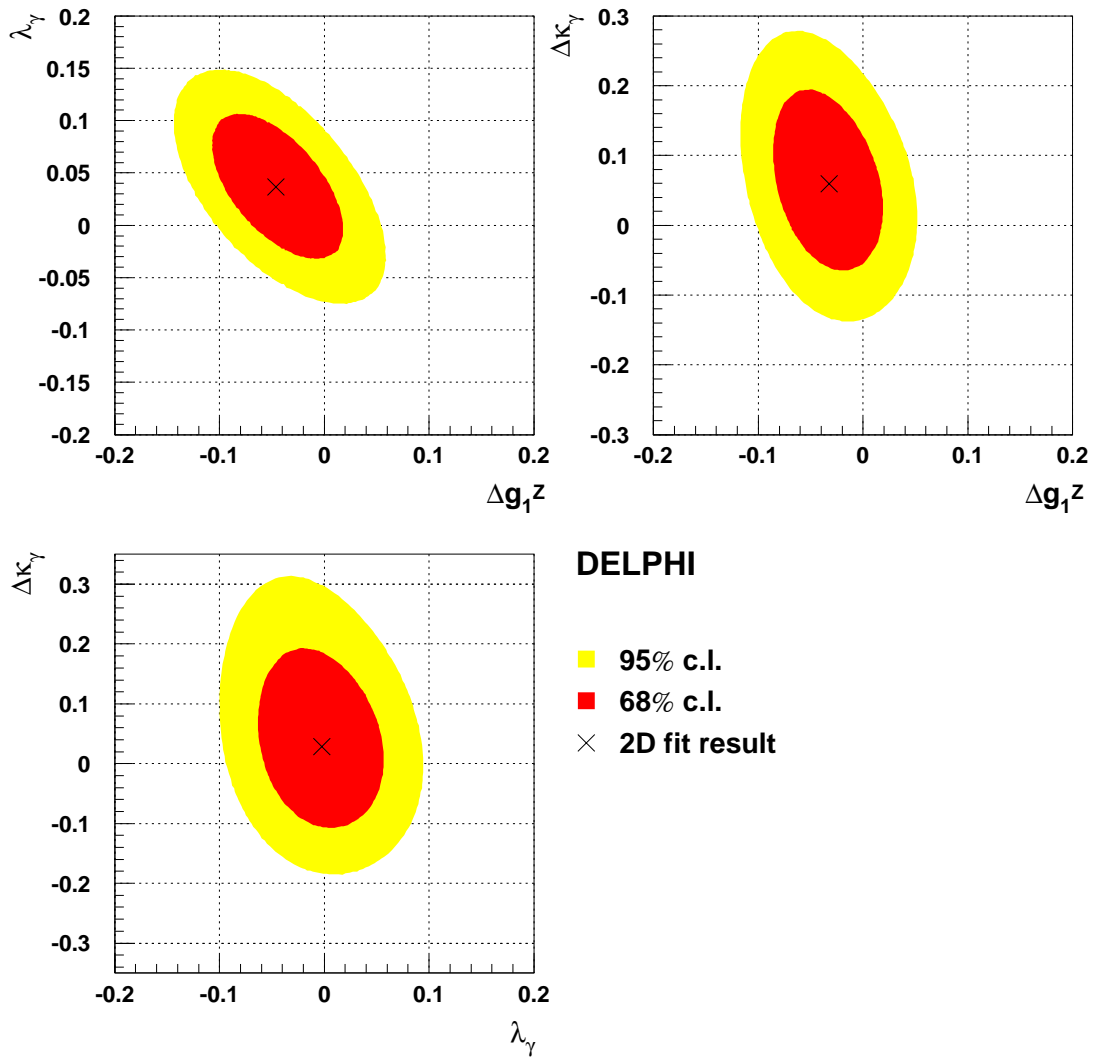,height=\textwidth}
\caption
    {The log-likelihood contours for two-parameter fits to the data, combining results from the semi-leptonic, fully hadronic and single $W$       final states. The plots include contributions from both statistical and systematic effects}
\label{fig_results_2d}
\end{figure}

\begin{figure}[!htbp]
\centering\epsfig{file=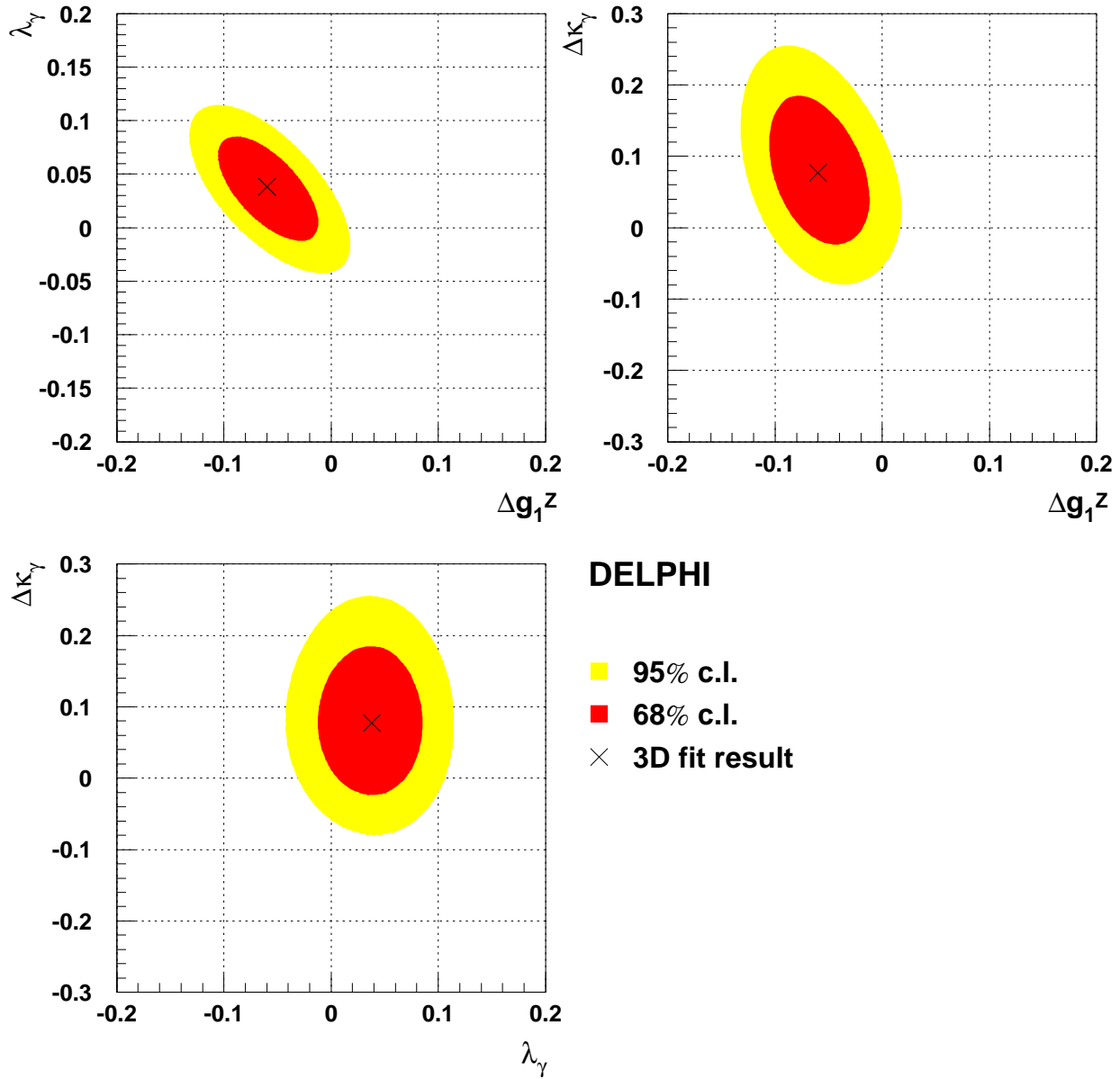,height=\textwidth}
\caption
{Intersections of the 68\% and 95\% confidence level 3-parameter log-likelihood surfaces with the three 2-parameter planes containing the minumum of the 3-parameter likelihood fit. The figures combine the results from the semi-leptonic, fully hadronic and single $W$ final states and include contributions from both statistical and systematic effects}
\label{fig_results_3d}
\end{figure}

The data taken by \DELPHI{} between 1998 and 2000 were collected at centre-of-mass energies between 189 and 209~GeV.  The results for the measurement of the couplings from single parameter fits to the data in the different channels are given in table~\ref{tab_results_1d} and the likelihood curves from these fits are shown in figure~\ref{fig_results_1d}.  The results of the simultaneous fits to the data for all combinations of two parameters (\dgz-\lgamma, \dgz-\dkg~and \lgamma-\dkg) are given in table~\ref{tab_results_2d}.  The corresponding likelihood contours are shown in figure~\ref{fig_results_2d}.  The result from the simultaneous fit to all three couplings is given in table~\ref{tab_results_3d} and the likelihood contours corresponding to the intersections of the three 2-parameter planes containing the minimum of the distribution with the three-dimensional 3-parameter likelihood distribution are shown in figure~\ref{fig_results_3d}. 

It may be noted from the results shown in tables~\ref{tab_results_1d}~-~\ref{tab_results_3d} that the 68\% and 95\% confidence levels obtained in the 3-parameter fit are somewhat narrower than those obtained in the 1-parameter fit  to the same parameter. This is not expected if the likelihood distributions are strictly Gaussian in form. However, such an effect is also observed in analysis of a significant fraction (5\%) of simulated event samples of the same size as the experimental sample.  As has been pointed out in previous studies of both simulated~\cite{SEKULIN} and experimental~\cite{RESULTS_OPAL} samples, the quadratic dependence of the cross-section on the couplings we consider does indeed lead to non-Gaussian likelihood distributions, which can thus explain this behaviour. The results for the multidimensional fits, in particular those for the 3-parameter fit, should therefore be viewed with this constraint on their interpretation in mind.

\begin{figure}[!htbp]
\begin{center}
\epsfig{file=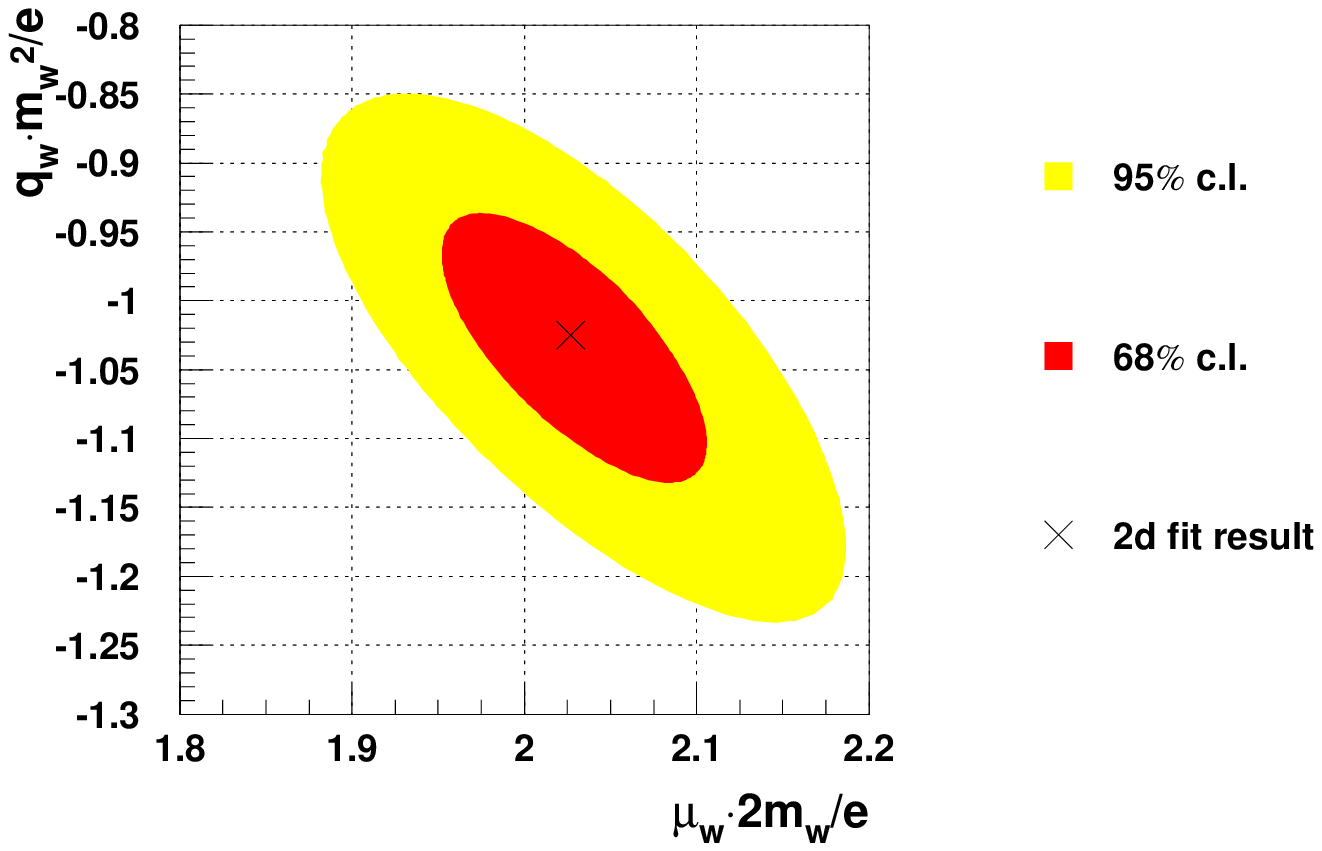,width=\textwidth}
\end{center}
\caption
    {The log-likelihood contours for a two-parameter fit to \qw\ and \muw, respectively the electric quadrupole and magnetic dipole moments of the $W^+$ boson, obtained from the simultaneous fit to \lgamma{} and \dkg. Results from the semi-leptonic, fully hadronic and single $W$ final states have been combined in the plot, and contributions from both statistical and systematic effects are included. The Standard Model expectations for the quantities plotted are: $ q_W m_W^2/e = -1$ and $\mu_W 2 m_W/e = 2$.}
\label{fig_results_multipole}
\end{figure}

The result from the simultaneous fit to \lgamma~and \dkg~can be converted to a measurement of the magnetic dipole moment, $\mu_W$, and the electric quadrupole moment, $q_W$, of the $W^+$ boson using the relations given in equations~(\ref{eqn_muw}) and~(\ref{eqn_qw}). The resulting two-parameter fit is shown in figure~\ref{fig_results_multipole}. The fitted values of $\mu_W$ and $q_W$ are

\begin{center}
  \begin{tabular}{rrcll}
              &  $\mu_W \cdot 2 m_W /e$ & = &  $\ \ 2.027_{-0.075}^{+0.078} $ & \\  
   and\    &  $q_W \cdot m_W^2 /e$      & = &  $ -1.025_{-0.088}^{+0.093} $ &  , \\
  \end{tabular}
\end{center}

\noindent where the errors include both statistical and systematic contributions. These results may be compared with the Standard Model predictions of $2$ and $-1$ for these two quantities, respectively.

\begin{table}[!htbp]
\begin{center}
\begin{tabular}{cccc}                                                  
\hline
\hline
Channel & \dgz  & \lgamma & \dkg  \\ 
\hline
\jjlv\  (Optimal Variables)                 
                          & $-0.024^{+0.033}_{-0.031}$ 
                          & \ \ $0.006^{+0.036}_{-0.035}$ 
                          & \ \ $0.014^{+0.103}_{-0.094}$ \\
\jjlv\ (Angular Variables)
                          & \ \ $0.006^{+0.040}_{-0.039}$ 
                          & \ \ $0.019^{+0.045}_{-0.043}$ 
                          & $-0.091^{+0.096}_{-0.085}$ \\
\jjjj                     & $-0.030^{+0.083}_{-0.067}$ 
                          & $-0.032^{+0.093}_{-0.070}$ 
                          & \ \ $0.031^{+0.196}_{-0.149}$ \\
single W                  & -- 
                          & \ \ $0.037^{+0.250}_{-0.288}$ 
                          & \ \ $0.027^{+0.131}_{-0.148}$ \\
\hline
\hline
Combined                  & $-0.025^{+0.033}_{-0.030} $ 
                          & \ \ $ 0.002^{+0.035}_{-0.035} $ 
                          & \ \ $ 0.024^{+0.077}_{-0.081} $ \\
\hline
\hline
\end{tabular}
\caption[]{The results for single parameter fits to the couplings in the individual channels. In each fit, the other two couplings were held at their Standard Model values. The errors given for the individual analyses are statistical; the systematic contributions are given in tables~\ref{tab_systematics_jjlv}, \ref{tab_systematics_jjjj} and \ref{tab_systematics_singleW}. As indicated in the text, the Angular Variables analysis of the \jjlv\ final state was performed as a cross-check, the values in the combination of all three channels being obtained using the results from the \jjlv\ Optimal Variables analysis. The combined results also contain the systematic errors, included via the combination method described in the text.} 
\label{tab_results_1d}
\end{center}
\end{table}
\begin{table}[htbp]
\begin{center}
\begin{tabular}{|l||r|r|rr|} \hline
Parameter   & \multicolumn{1}{|c|}{68\% C.L.}  & \multicolumn{1}{|c|}{95\% C.L.}   & \multicolumn{2}{|c|}{Correlations}  \\
\hline \hline
\dgz       & $-0.046^{+0.040}_{-0.040}$ & [$-0.123,~~+0.035$] & $1.0$  & $-0.49$  \\  
\lgamma   & $0.037^{+0.045}_{-0.044}$ & [$-0.051,~~+0.124$]   & $-0.49$  & $1.0$ \\
\hline
\dgz       & $-0.033^{+0.032}_{-0.033}$ & [$-0.097,~~+0.032$]  & $1.0$ & $-0.41$ \\  
\dkg       & $0.059^{+0.088}_{-0.079}$ & [$-0.093,~~+0.233$]    & $-0.41$ & $1.0$ \\
\hline
\lgamma        & $-0.002^{+0.035}_{-0.035}$ & [$-0.070,~~+0.067$]  & $1.0$ & $0.10$ \\
\dkg        & $0.028^{+0.083}_{-0.077}$ & [$-0.120,~~+0.198$]  & $0.10$  & $1.0$ \\ 
\hline
\end{tabular}
\caption{ The measured central values, one standard deviation errors and
    limits at 95\% confidence level obtained by combining the different
    channels in the 3 two-parameter fits.  Since the shape of the
    log-likelihood is not parabolic, there is some ambiguity in the
    definition of the correlation coefficients and the values quoted here are
    approximate.  In each fit, the listed parameters were varied while the remaining one was
    fixed to its Standard Model value.  Both statistical and systematic
    errors are included.  Note that the 68\% and 95\% confidence  limits reported here refer to single-parameter errors (in contrast to those shown in the two-parameter plots of figure~\ref{fig_results_2d})
    and are defined by  $\Delta\ln {\mathcal L}=+0.5$ and $\Delta\ln{\mathcal L} =+1.92$, respectively.}
\label{tab_results_2d}
\end{center}
\end{table}
\begin{table}[htbp]
\begin{center}
\begin{tabular}{|l||r|r|rrr|} \hline
Parameter & \multicolumn{1}{|c|}{68\% C.L.}  & \multicolumn{1}{|c|}{95\% C.L.} & \multicolumn{3}{|c|}{Correlations} \\
                     &             &            & \dgz & \lgamma & \dkg \\
\hline \hline
\dgz              & $-0.060^{+0.031}_{-0.030}$ & [$-0.118,~~+0.002$]   & $1.0$ & $-0.55$ & $-0.41$ \\  
\lgamma       & $0.038^{+0.031}_{-0.032}$ & [$-0.027,~~+0.099$]  & $-0.55$ & $1.0$ & $-0.04$ \\
\dkg               & $0.077^{+0.070}_{-0.070}$ & [$-0.050,~~+0.218$]     & $-0.41$ & $-0.04$ & $1.0$ \\ 
\hline
\end{tabular}
\caption{The measured central values, one standard deviation errors and
    limits at 95\% confidence level, obtained by combining the different
    channels in the three-parameter fit.  Since the shape of the
    log-likelihood is not parabolic, there is some ambiguity in the
    definition of the correlation coefficients and the values quoted here are
    approximate.  Both statistical and systematic errors are included.  Note that the 
    68\% and 95\% confidence  limits reported refer to single-parameter errors and are defined by  $\Delta\ln {\mathcal L}=+0.5$ and $\Delta\ln{\mathcal L} =+1.92$, respectively.}
\label{tab_results_3d}
\end{center}
\end{table}

\section{Conclusions}
\label{sec_conclusions}

The data taken by \DELPHI{} at centre-of-mass energies between 189~and 209~GeV have been used to probe the non-Abelian nature of the Standard Model. Limits have been placed on the trilinear gauge boson couplings which describe the $WWZ$ and $WW\gamma$ vertices; in particular, reactions leading to $W$ pair production and single $W$ production have been used to set limits on the parameters \dgz, \lgamma\ and \dkg.   The combined results for fits to a single parameter, where the other two parameters were held at their Standard Model values, are:
 
\begin{center}
  \begin{tabular}{rlcrl}
              &  --0.084 $\mathrm{<}$ & \dgz{}    & $\mathrm{<}$ 0.039  &,\\
              &  --0.065 $\mathrm{<}$ & \lgamma{} & $\mathrm{<}$ 0.071 &, \\
    and\   & --0.129 $\mathrm{<}$ & \dkg{}    & $\mathrm{<}$ 0.182 & \\
  \end{tabular}
\end{center}

\noindent at 95\% confidence level.  Fits were also made where two or three parameters were allowed to vary simultaneously.  No deviations from the Standard Model predictions have been observed.

\subsection*{Acknowledgements}
\vskip 3 mm
We are greatly indebted to our technical 
collaborators, to the members of the CERN-SL Division for the excellent 
performance of the LEP collider, and to the funding agencies for their
support in building and operating the DELPHI detector.\\
We acknowledge in particular the support of \\
Austrian Federal Ministry of Education, Science and Culture,
GZ 616.364/2-III/2a/98, \\
FNRS--FWO, Flanders Institute to encourage scientific and technological 
research in the industry (IWT) and Belgian Federal Office for Scientific, 
Technical and Cultural affairs (OSTC), Belgium, \\
FINEP, CNPq, CAPES, FUJB and FAPERJ, Brazil, \\
Ministry of Education of the Czech Republic, project LC527, \\
Academy of Sciences of the Czech Republic, project AV0Z10100502, \\
Commission of the European Communities (DG XII), \\
Direction des Sciences de la Mati$\grave{\mbox{\rm e}}$re, CEA, France, \\
Bundesministerium f$\ddot{\mbox{\rm u}}$r Bildung, Wissenschaft, Forschung 
und Technologie, Germany,\\
General Secretariat for Research and Technology, Greece, \\
National Science Foundation (NWO) and Foundation for Research on Matter (FOM),
The Netherlands, \\
Norwegian Research Council,  \\
State Committee for Scientific Research, Poland, SPUB-M/CERN/PO3/DZ296/2000,
SPUB-M/CERN/PO3/DZ297/2000, 2P03B 104 19 and 2P03B 69 23(2002-2004),\\
FCT - Funda\c{c}\~ao para a Ci\^encia e Tecnologia, Portugal, \\
Vedecka grantova agentura MS SR, Slovakia, Nr. 95/5195/134, \\
Ministry of Science and Technology of the Republic of Slovenia, \\
CICYT, Spain, AEN99-0950 and AEN99-0761,  \\
The Swedish Research Council,      \\
The Science and Technology Facilities Council, UK, \\
Department of Energy, USA, DE-FG02-01ER41155, \\
EEC RTN contract HPRN-CT-00292-2002. \\


\newpage


\newpage

%
%
%

%


\begin{thebibliography}{99}                              
                       
\bibliographystyle{aip}

\bibitem{YELTGC}
{G. Gounaris, J.-L. Kneur and D. Zeppenfeld},
\newblock {\it Triple Gauge Boson Couplings},
\newblock in {\em Physics at LEP2 Vol. 1}, edited by {G. Altarelli, T. Sj\"ostrand and F. Zwirner}, 
\newblock CERN 96-01 (1996).

\bibitem{HAGIWARA}
{K. Hagiwara, R. Peccei, D. Zeppenfeld and K. Hikasa},
\newblock Nucl. Phys. {\bf B282} (1987) 253.

\bibitem{DELPHI_TGC172}
{DELPHI Collaboration, P. Abreu {\it et al.}},
\newblock Phys. Lett. {\bf B423} (1998) 194. 

\bibitem{DELPHI_TGC183}
{DELPHI Collaboration, P. Abreu {\it et al.}},
\newblock Phys. Lett. {\bf B459} (1999) 382.

\bibitem{DELPHI_TGC189}
{DELPHI Collaboration, P. Abreu {\it et al.}},
\newblock Phys. Lett. {\bf B502} (2001) 9.

\bibitem{DELPHI_SDM}
{DELPHI Collaboration, J. Abdallah {\it et al.}},
\newblock Eur. Phys. J. {\bf C54} (2008) 345.

\bibitem{RESULTS_ALEPH}
{ALEPH Collaboration, S. Schael {\it et al.}},
\newblock Phys. Lett. {\bf B614} (2005) 7.

\bibitem{RESULTS_L3}
{L3 Collaboration, P. Achard {\it et al.}},
\newblock Phys. Lett. {\bf B586} (2004) 151.

\bibitem{RES_SINGLEW_L3}
{L3 Collaboration, P. Achard {\it et al.}}, 
\newblock Phys. Lett. {\bf B547} (2002) 151.

\bibitem{RESULTS_OPAL}
{OPAL Collaboration, G. Abbiendi {\it et al.}},
\newblock Eur. Phys. J. {\bf C33} (2004) 463. 


\bibitem{DELPHI}
{DELPHI Collaboration, P. Aarnio {\it et al.}},
\newblock Nucl. Instr. Meth. {\bf A303} (1991) 233.

\bibitem{DELPHI_PERFORMANCE}
{DELPHI Collaboration, P. Abreu {\it et al.}},
\newblock Nucl. Instr. Meth. {\bf A378} (1996) 57.

\bibitem{TRIGGER}
{DELPHI Trigger Group, A. Augustinus {\it et al.}},
\newblock Nucl. Instr. Meth. {\bf A515} (2003) 782.

\bibitem{SITRACKER}
{DELPHI Silicon Tracker Group, P. Chochula {\it et al.}},
\newblock Nucl. Instr. Meth. {\bf A412} (1998) 304.

\bibitem{LUMI}
{S.J. Alvsvaag {\it et al.}},
\newblock Nucl. Instr. Meth. {\bf A425} (1999) 106.

\bibitem{WPHACT1}
{E. Accomando and A. Ballestrero}, 
\newblock Comp. Phys. Comm. {\bf 99} (1997) 270.

\bibitem{WPHACT2}
{E. Accomando, A. Ballestrero and E. Maina}, 
\newblock Comp. Phys. Comm. {\bf 150} (2003) 166.

\bibitem{WPHAC_SETUP}
{A. Ballestrero, R. Chierici, F. Cossutti and E. Migliore}, 
\newblock Comp. Phys. Comm. {\bf 152} (2003) 175.

\bibitem{YFSWW}
{S. Jadach, W. Placzek, M. Skrzypek, B.F.L. Ward and Z. Was}, 
\newblock Comp. Phys. Comm. {\bf 140} (2001) 432.

\bibitem{KK2f}
{S. Jadach, B.F.L. Ward and Z. Was}, 
\newblock Comp. Phys. Comm. {\bf 130} (2000) 260.

\bibitem{KoralZ}
{S. Jadach, B.F.L. Ward and Z. Was}, 
\newblock Comp. Phys. Comm. {\bf 79} (1994) 503.

\bibitem{BDK}
{F.A. Berends, P.H. Daverveldt and R. Kleiss},
\newblock Comp. Phys. Comm. {\bf 40} (1986) 285.

\bibitem{BDKRC}
{F.A. Berends, P.H. Daverveldt and R. Kleiss},
\newblock Comp. Phys. Comm. {\bf 40} (1986) 271.

\bibitem{PYTHIA1}
{T. Sj\"ostrand},
\newblock {Comp. Phys. Comm.} {\bf 82} (1994) 74.

\bibitem{PYTHIA2}
{T. Sj\"ostrand},
\newblock {\it PYTHIA 5.719 / JETSET 7.4},
\newblock in {\em Physics at LEP2 Vol. 1}, edited by {G. Altarelli, T. Sj\"ostrand and F. Zwirner}, 
\newblock CERN 96-01 (1996).

\bibitem{TUNING}
{DELPHI Collaboration, P. Abreu {\it et al.}}, 
\newblock Zeit. Phys. {\bf C73} (1996) 11.

\bibitem{WWXSEC}
{DELPHI Collaboration, J. Abdallah {\it et al.}}, 
\newblock Eur. Phys. J. {\bf C34} (2004) 127.

\bibitem{IDA1}
{T.G.M. Malmgren},
\newblock Comp. Phys. Comm. {\bf 28} (1983) 229.

\bibitem{IDA2}
{T.G.M. Malmgren and K.E. Johansson},
\newblock Nucl. Instr. Meth. {\bf A403} (1998) 481.

\bibitem{DURHAM}
{S. Catani {\it et al.}},
\newblock Phys. Lett. {\bf B269} (1991) 432.

\bibitem{SPRIME}
{P. Abreu {\it et al.}},
\newblock Nucl. Instr. Meth. {\bf A427} (1999) 487.

\bibitem{JETNET}
{C. Peterson, T. R\"ognvaldsson and L. L\"onnblad},
\newblock {Comp. Phys. Comm.} {\bf 81} (1994) 185.

\bibitem{NNOPTIMISATION}
{K.-H. Becks, J. Drees, U. Flagmeyer and U. M\"uller},
\newblock Nucl. Instr. Meth. {\bf A426} (1999) 599.

\bibitem{FOX-WOLFRAM}
{G.C. Fox and S. Wolfram}, 
\newblock Nucl. Phys. {\bf B149} (1979) 413. 

\bibitem{DELTGC}
{O.P. Yushchenko and V.V. Kostyukhin}, 
\newblock {\it DELTGC -- A program for four-fermion calculations}, 
\newblock {DELPHI note DELPHI 99-4 PHYS 816,} \\
\newblock {http://delphiwww.cern.ch/pubxx/delnote/public/99\_04\_phys\_816.ps.gz (1999).}

\newpage

\bibitem{OO1}
{M. Diehl and O. Nachtmann},
\newblock Z. Phys. {\bf C62} (1994) 397.

\bibitem{OO2}
{C. Papadopoulos},
\newblock Phys. Lett. {\bf B386} (1996) 442.

\bibitem{OO3}
{M. Diehl and O. Nachtmann},
\newblock Eur. Phys. J. {\bf C1} (1998) 177.

\bibitem{OOTGC}
{G.K. Fanourakis, D.A. Fassouliotis and S.E. Tzamarias},
\newblock Nucl. Instr. Meth. {\bf A414} (1999) 399.

\bibitem{OOOPTIMISATION}
{G.K. Fanourakis, D.A. Fassouliotis, A. Leisos, N. Mastroyiannopoulos and S.E. Tzamarias}, 
\newblock Nucl. Instr. Meth. {\bf A430} (1999) 474.

\bibitem{BINNING}
{G.K. Fanourakis, D.A. Fassouliotis, A. Leisos, N. Mastroyiannopoulos and S.E. Tzamarias}, 
\newblock Nucl. Instr. Meth. {\bf A430} (1999) 455.

\bibitem{HERWIG}
{G. Corcella {\it et al.}},
\newblock Comp. Phys. Comm. {\bf 67} (1992) 465.

\bibitem{ARIADNE}
{L. L\"onnblad},
\newblock Comp. Phys. Comm. {\bf 71} (1992) 15.

\bibitem{ALEPHjjjj}
{R. Barate {\it et al.}},
\newblock Phys. Lett. {\bf B422} (1998) 369.

\bibitem{LEP2MCWorkshop}
{M. Gr\"unewald {\it et al.}},
\newblock {\it Four Fermion Production in Electron-Positron Collisions},
\newblock in {\em Reports of the Working Groups on Precision Calculations for
LEP2 Physics}, edited by {S. Jadach, G. Passarino and R. Pittau}, 
\newblock CERN 2000-009 (2000).

\bibitem{Gentle2}
{D. Bardin {\it et al.}},
\newblock Comp. Phys. Comm. {\bf 104} (1997) 161.

\bibitem{RacoonWW}
{A. Denner {\it et al.}},
\newblock Phys. Lett. {\bf B475} (2000) 127.

\bibitem{RAD_CORR}
{R. Chierici and F. Cossutti},
\newblock Eur. Phys. J. {\bf C9} (1999) 449.

\bibitem{LEP_BEAM}
{The LEP Energy Working Group, R. Assmann {\it et al.}},
\newblock Eur. Phys. J. {\bf C39} (2005) 253.

\bibitem{BHABHA}
{W. Placzek {\it et al.}},
\newblock {\it Precision Calculation of Bhabha Scattering at LEP}, CERN-TH 99-07,
hep-ph/9903381 (1999).

\bibitem{MLBZ}
{DELPHI Collaboration, J. Abdallah {\it et al.}}, 
\newblock Eur.~Phys.~J. {\bf C55} (2008) 1.

\bibitem{SK1}
{T. Sj\"ostrand and V.A. Khoze},
\newblock Z. Phys. {\bf C62} (1994) 281.

\bibitem{CR}
{DELPHI Collaboration, J. Abdallah {\it et al.}},
\newblock Eur. Phys. J. {\bf C51} (2007) 249.

\bibitem{L3CR}
{L3 Collaboration,  P.~Achard {\it et al.}}, 
\newblock Phys. Lett. {\bf B561} (2003) 202. 

\bibitem{OPALCR}
{OPAL Collaboration, G.~Abbiendi {\it et al.}},
\newblock Eur. Phys. J. {\bf C45} (2006) 291. 

\bibitem {ALEPHCR}
{ALEPH Collaboration, S.~Schael {\it et al.}},
\newblock Eur. Phys. J. {\bf C47} (2006) 309.

\bibitem{LUBOEI}
{L. L\"onnblad and T. Sj\"ostrand},
\newblock Eur. Phys. J. {\bf C2} (1998) 165.

\bibitem{DBEC}
{DELPHI Collaboration, J. ~Abdallah {\it et al.}},
\newblock Eur. Phys. J. {\bf C44} (2005) 161.  

\bibitem{L3BEC}
{L3 Collaboration, P. Achard {\it et al.}},
\newblock Phys. Lett. {\bf B547} (2002) 139. 

\bibitem{OPALBEC}
{OPAL Collaboration, G.~Abbiendi {\it et al.}},
\newblock Eur. Phys. J. {\bf C36} (2004) 297. 

\bibitem{ALEPHBEC}
{ALEPH Collaboration, S.~Schael {\it et al.}},
\newblock Phys. Lett. {\bf B606} (2005) 265. 


\bibitem{BORUT}
{B.P. Ker\v{s}evan, B. Golob, G. Kernel and T. Podobnik},
\newblock {\it Estimation of Confidence Intervals in Measurements of Trilinear Gauge Boson Couplings}, 
\newblock in {\em Proceedings of the Workshop on Confidence Limits}, edited by {L. Lyons and F. James}, 
\newblock CERN 2000-05 (2000).

\bibitem{SEKULIN}
{R. L. Sekulin},
\newblock Phys. Lett. {\bf B338} (1994) 369.


\end{thebibliography}
\end{document}